\documentclass[%
aps,
prl,
reprint,
superscriptaddress,
nofootinbib,
amsmath,
amssymb
]{revtex4-2}
\usepackage{
	physics,
	graphicx,
	multirow,
	tabularx,
	mathtools, %for \begin{rcases} \end{rcases}
	placeins, %To include \FloatBarrier
	nicefrac, %For inline fraction
	hyperref, %To use put in url, enable links in bibiliography
	threeparttable, % For table footnotes
	siunitx,
	enumitem,
	makecell,
	amsmath,amssymb,amsthm,mathrsfs,amsfonts,dsfont
}
\usepackage{soul}
\usepackage[capitalize]{cleveref}
\Crefname{section}{Sec.}{Secs.}

\usepackage[dvipsnames]{xcolor}
\hypersetup{
	colorlinks,
	linkcolor={blue!50!black},
	citecolor={blue!50!black},
	urlcolor={blue!80!black}
}% remove the box around citation link and bibilio link

\usepackage[caption=false]{subfig} %The caption part is not compatible with revtex
\begin{document}

%\title{Experimental Channel Purification in Optical Quantum Network}
\title{Experimental Quantum Channel Purification}

\author{Yue-Yang Fei}
\altaffiliation{These authors contributed equally to this work.}
\affiliation{Hefei National Research Center for Physical Sciences at the Microscale and School of Physical Sciences, University of Science and Technology of China, Hefei 230026, China}
\affiliation{Shanghai Research Center for Quantum Science and CAS Center for Excellence in Quantum Information and Quantum Physics, University of Science and Technology of China, Shanghai 201315, China}

\author{Zhenhuan Liu}
\altaffiliation{These authors contributed equally to this work.}
\affiliation{Center for Quantum Information, Institute for Interdisciplinary Information Sciences, Tsinghua University, Beijing 100084, China}

\author{Rui Zhang}
\affiliation{Hefei National Research Center for Physical Sciences at the Microscale and School of Physical Sciences, University of Science and Technology of China, Hefei 230026, China}
\affiliation{Shanghai Research Center for Quantum Science and CAS Center for Excellence in Quantum Information and Quantum Physics, University of Science and Technology of China, Shanghai 201315, China}

\author{Zhenyu Cai}
\affiliation{Department of Materials, University of Oxford, Parks Road, Oxford OX1 3PH, United Kingdom}

\author{Xu-Fei Yin}
\affiliation{Hefei National Research Center for Physical Sciences at the Microscale and School of Physical Sciences, University of Science and Technology of China, Hefei 230026, China}
\affiliation{Shanghai Research Center for Quantum Science and CAS Center for Excellence in Quantum Information and Quantum Physics, University of Science and Technology of China, Shanghai 201315, China}
%\affiliation{Hefei National Laboratory, University of Science and Technology of China, Hefei 230088, China}

\author{Yingqiu Mao}
\affiliation{Hefei National Research Center for Physical Sciences at the Microscale and School of Physical Sciences, University of Science and Technology of China, Hefei 230026, China}
\affiliation{Shanghai Research Center for Quantum Science and CAS Center for Excellence in Quantum Information and Quantum Physics, University of Science and Technology of China, Shanghai 201315, China}
%\affiliation{Hefei National Laboratory, University of Science and Technology of China, Hefei 230088, China}

\author{Li Li}
\affiliation{Hefei National Research Center for Physical Sciences at the Microscale and School of Physical Sciences, University of Science and Technology of China, Hefei 230026, China}
\affiliation{Shanghai Research Center for Quantum Science and CAS Center for Excellence in Quantum Information and Quantum Physics, University of Science and Technology of China, Shanghai 201315, China}
\affiliation{Hefei National Laboratory, University of Science and Technology of China, Hefei 230088, China}

\author{Nai-Le Liu}
\affiliation{Hefei National Research Center for Physical Sciences at the Microscale and School of Physical Sciences, University of Science and Technology of China, Hefei 230026, China}
\affiliation{Shanghai Research Center for Quantum Science and CAS Center for Excellence in Quantum Information and Quantum Physics, University of Science and Technology of China, Shanghai 201315, China}
\affiliation{Hefei National Laboratory, University of Science and Technology of China, Hefei 230088, China}

\author{Yu-Ao Chen}
%\email{yuaochen@ustc.edu.cn}
\affiliation{Hefei National Research Center for Physical Sciences at the Microscale and School of Physical Sciences, University of Science and Technology of China, Hefei 230026, China}
\affiliation{Shanghai Research Center for Quantum Science and CAS Center for Excellence in Quantum Information and Quantum Physics, University of Science and Technology of China, Shanghai 201315, China}
\affiliation{Hefei National Laboratory, University of Science and Technology of China, Hefei 230088, China}
\affiliation{New Cornerstone Science Laboratory, School of Emergent Technology, University of Science and Technology of China, Hefei 230026, China}

\author{Jian-Wei Pan}
\affiliation{Hefei National Research Center for Physical Sciences at the Microscale and School of Physical Sciences, University of Science and Technology of China, Hefei 230026, China}
\affiliation{Shanghai Research Center for Quantum Science and CAS Center for Excellence in Quantum Information and Quantum Physics, University of Science and Technology of China, Shanghai 201315, China}
\affiliation{Hefei National Laboratory, University of Science and Technology of China, Hefei 230088, China}

\begin{abstract}
Quantum networks, which integrate multiple quantum computers and the channels connecting them, are crucial for distributed quantum information processing but remain inherently susceptible to channel noise. 
Channel purification emerges as a promising technique for suppressing noise in quantum channels without complex encoding and decoding operations, making it particularly suitable for remote quantum information transmission in optical systems.
In this work, we introduce an experimental setup for efficient channel purification, harnessing the spatial and polarization properties of photons. 
Our design employs two Fredkin gates to enable coherent interference between independent noise channels, achieving effective noise suppression across a wide range of noise levels and types.
Through application to entanglement distribution, our protocol demonstrates a superior capability to preserve entanglement against channel noise compared to conventional entanglement purification methods.
\end{abstract}

\maketitle

\textbf{Introduction.}
Quantum networks serve as the fundamental infrastructure that integrates the capabilities of various quantum devices, playing a crucial role in quantum cryptography and multi-party quantum computation~\cite{Simon2017network,chen2021integrated,main2025distributed}.
Photons are the preferred quantum information carriers for inter-node communication due to their superior transmission characteristics~\cite{pan2012multiphoton}. 
However, the scalability of quantum networks is fundamentally limited by noise accumulation within quantum channels, hindering reliable long-distance quantum information transmission.

\begin{figure}[htbp!]
\centering
\includegraphics[width=1\linewidth]{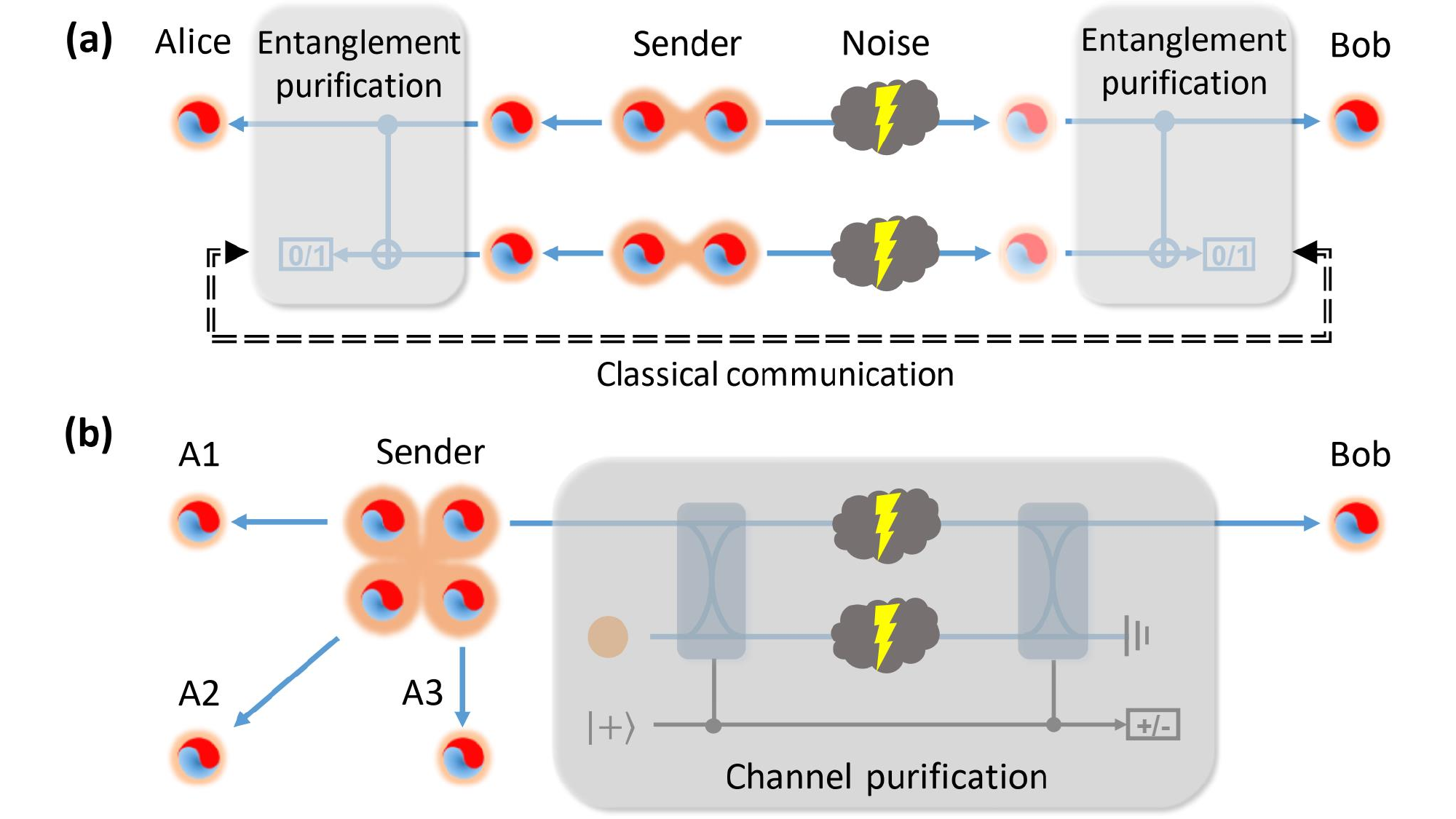}
\caption{Difference between entanglement purification and channel purification.
(a) In a state distribution task, as demonstrated with a bipartite one, even if the noise is primarily introduced by a single channel (from the Sender to Bob), entanglement purification requires operations over all parties, leading to high resource consumption.
(b) A channel purification protocol specifically targets the noise channel and suppresses noise with only local operations.
The protocol does not require operations on the other channels from the sender to A$_1$, A$_2$, or A$_3$.
Furthermore, channel purification can operate on a single copy of the distributed state, and has no requirement for quantum memory to store multiple noisy states.}
\label{fig:protocol}
\end{figure}

Addressing noise in quantum networks primarily involves quantum error correction~\cite{Devitt2013qec,terhal2015memory} and entanglement purification~\cite{bennett1996purification,Pan2001purification}. 
Quantum error correction, while experimentally validated~\cite{Ni2023breakeven,Bluvstein2024qec,Acharya2025surface}, proves challenging for optical implementation due to its requirement for system scaling and complex encoding/decoding operations. 
Entanglement purification, relying on local operations and classical communication, distills high-fidelity entanglement from multiple noisy states.
As illustrated in Fig.~\ref{fig:protocol}(a), these operations are applied to the entire state rather than specifically targeting the qubit passing through the noisiest channel.
Furthermore, entanglement purification protocols often necessitate quantum memory for state storage, adding substantial implementation complexity.

Channel purification~\cite{liu2024virtualchannelpurification,PhysRevLett.131.230601,PhysRevA.110.052431} provides a distinct alternative to quantum error correction and entanglement purification. 
This protocol leverages multiple noisy channels to effectively obtain a channel with a reduced noise rate, drawing inspiration from state purification~\cite{cirac1999optimal,cotler2019cooling,hugginsVirtualDistillationQuantum2021,koczorExponentialErrorSuppression2021} and coherent control~\cite{chiribella2019quantum}. 
Unlike its counterparts, channel purification targets the noisiest channel without complex encoding/decoding or quantum memory requirements, as illustrated in Fig.~\ref{fig:protocol}(b). 
Current approaches rely on implementing multiple Fredkin gates, a computationally powerful three-qubit gate that, despite its broad utility in quantum error mitigation~\cite{hugginsVirtualDistillationQuantum2021,koczorExponentialErrorSuppression2021}, quantum simulation~\cite{Lloyd2014qpca,cotler2019cooling}, and quantum property testing~\cite{Zhou2024hybrid,faehrmann2025hadamard}, poses significant experimental challenges.

In this work, we design a novel experimental configuration on a linear optical platform to efficiently implement multiple Fredkin gates in a single circuit by exploiting interactions between the spatial and polarization degrees of freedom (DoF) of photons.
Using this configuration, we experimentally demonstrate a channel purification protocol and apply it to various noise channels.
Through process tomography and average fidelity estimation, we show that this protocol significantly reduces noise rates, confirming its effectiveness and universality.
Furthermore, we showcase the practical advantage of our protocol by applying it to an entanglement distribution task, where it surpasses entanglement purification in preserving distributed entanglement.

\textbf{Channel purification.}
The target of channel purification is to consume several copies of the noise channel $\mathcal{C}$ and obtain a less noisy one $\mathcal{C}_{\mathrm{p}}$.
Given a target channel $\mathcal{U}$, which is typically a unitary channel, the only requirement of a channel purification protocol is that, for any input state $\rho$, the output of purified channel $\mathcal{C}_{\mathrm{p}}(\rho)$ is closer to $\mathcal{U}(\rho)$ compared with $\mathcal{C}(\rho)$.
Note that, according to this definition, a quantum error correction code can also be viewed as a channel purification protocol.
However, traditional quantum error correction aims to eliminate specific error components through complex encoding/decoding and numerous ancillary qubits~\cite{Devitt2013qec,terhal2015memory}. 
By relaxing these requirements, simpler channel purification protocols can be developed, potentially offering a more practical approach for linear optical quantum computing platforms.

\begin{figure*}[htbp!]
	\centering
	\includegraphics[width=1 \linewidth]{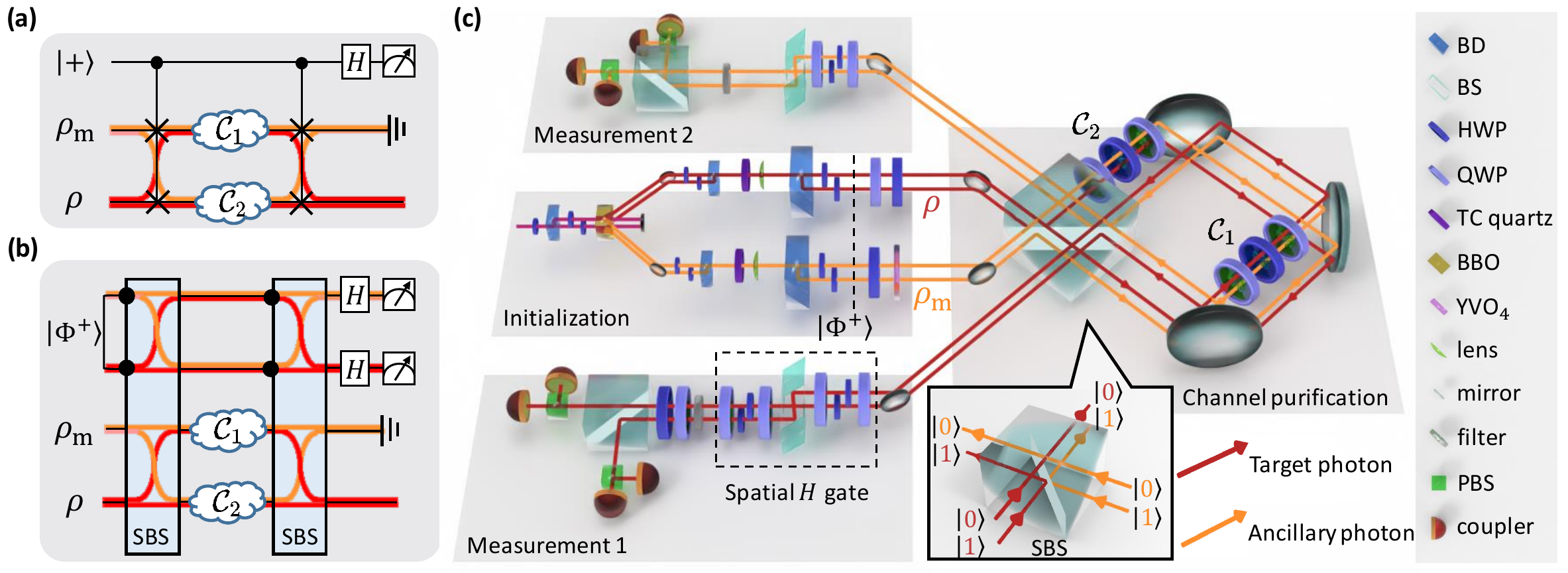}
\caption{Channel purification protocol and experimental setup.
(a) Quantum circuit of our channel purification protocol, where $H$ represents the Hadamard gate and $\rho_{\mathrm{m}}$ represents the maximally mixed state.
Two noise channels, $\mathcal{C}_1$ and $\mathcal{C}_2$, are sandwiched by two Fredkin gates.
(b) The circuit implemented in the linear optical system.
A Fredkin-like gate is realized by the spatial beam splitter (SBS).
Control register encoded in spatial DoF directs the photon to different paths.
Qubits in different DoF from the same photon are marked as the same color.
(c) Detailed experiment setup.
In order to distinguish two photons, we mark photons of the main and ancillary registers with red and orange colors, respectively.
The separation between these two light paths is 3 $\mathrm{mm}$.
QWP-HWP-QWP combinations are employed to compensate for phase shifts introduced by BS.
A total of 8 superconducting nanowire single-photon detectors are used for detection.
BD: beam displacer; BS: beam splitter; HWP: half wave plate; QWP: quarter wave plate; TC quartz: quartz for time compensation; PBS: polarizing beam splitter.
} 
\label{fig:setup}
\end{figure*}

We employ the channel purification protocol introduced in Ref.~\cite{liu2024virtualchannelpurification} with the circuit shown in Fig.~\ref{fig:setup}(a). 
This circuit mainly consists of three registers: the main register initialized in the input state $\rho$, the ancillary register in the maximally mixed state $\rho_{\mathrm{m}}$, and the control register in the state $\ket{+}=(\ket{0}+\ket{1})/{\sqrt{2}}$.
After initialization, one sequentially performs the Fredkin gate, two noise channels, and another Fredkin gate.
The Fredkin gate, also known as the controlled-SWAP gate, exchanges the main and ancillary registers based on the state of the control qubit.
At the end of the circuit, one measures the control qubit in Pauli-$X$ basis, discards the ancillary register and keeps the main register if the measurement result of the control qubit is $\ket{+}$.
With the input state $\rho$ and the post-selected output state on the main register, we obtain the purified channel $\mathcal{C}_{\mathrm{p}}$.

This channel purification protocol has many advantages.
Firstly, it has minimal requirements of the form of noise channels. 
Using Pauli twirling, every noise channel can be transformed into a Pauli-diagonal channel, $\mathcal{C}(\rho)=\sum_ip_iE_i\rho E_i^\dagger$ with $\{E_i\}_i$ being Pauli matrices, $E_0$ being the identity matrix, and $\{p_i\}_i$ being probabilities~\cite{wallmanNoiseTailoringScalable2016,caiConstructingSmallerPauli2019,tsubouchi2024symmetric}. 
When two noise channels are identical, the purified channel has the form of $\mathcal{C}_{\mathrm{p}}(\rho)=\sum_ip_i^\prime E_i\rho E_i^\dagger$ with
\begin{equation}\label{eq:cp}
p_i^\prime=p_i\frac{1+p_i}{1+\sum_jp_j^2}.
\end{equation}
It can be easily verified that, if $p_0$ is the largest probability in the original noise channel, $p_0^\prime\ge p_0$.
Therefore, in practical cases when the noise is not dominant, making the target channel $\mathcal{C}_0(\rho)=E_0\rho E_0^\dagger$ to be the principal component of $\mathcal{C}$, the purified channel is always closer to the target channel.
Moreover, the two input channels are not necessarily equivalent or single-qubit.
The only requirement for two noise channels is that their principal components after Pauli twirling are all the noiseless channel.

Although this protocol is built on the post-selection of the control qubit measurement result, the discarded data can also be utilized~\cite{faehrmann2025hadamard}.
One can evaluate the difference between channels induced by $\ket{+}$ and $\ket{-}$ measurement results on the control qubit and get a virtual channel $\mathcal{C}_{\mathrm{vp}}(\rho)=\sum_ip_i^{\prime\prime}E_i\rho E_i^\dagger$ with
\begin{equation}\label{eq:vcp}
p_i^{\prime\prime}=\frac{p_i^2}{\sum_jp_j^2}.
\end{equation}
While $\mathcal{C}_{\mathrm{vp}}$ is a non-physical channel limited to expectation value evaluation, it offers a methodology for more effectively analyzing measurement data and suppressing noise rate as $p_0^{\prime\prime}\ge p_0^\prime$~\cite{caiQuantumErrorMitigation2023,yuan2024virtual}.
We leave the detailed circuit analysis and derivations of Eq.~\eqref{eq:cp} and Eq.~\eqref{eq:vcp} in Supplementary Material~\cite{SM}.

\textbf{Experiment.}
The successful implementation of our channel purification protocol critically depends on the reliable realization of sequential Fredkin gates before and after two independent noise channels.
Current linear-optical implementations face significant limitations for realizing our protocol. 
Post-selection-based approaches impede deterministic multi-Fredkin gate execution~\cite{raj2016fredkin,Ono2017-ag}; while spatial encoding introduces correlated noise, hindering practical quantum circuit implementation~\cite{peng2025fredkin}. 
These drawbacks necessitate the exploration and development of alternative architectural designs.

Our new architecture is sketched in Fig.~\ref{fig:setup}(b).
Two photons, distinguished by different colors, are utilized to encode four qubits with polarization and spatial DoF.
The main register $\rho$ and the ancillary register $\rho_{\mathrm{m}}$ are prepared in the polarization DoF.
The two spatial qubits are prepared in the Bell state $\ket{\Phi^+}=(\ket{00}+\ket{11})/{\sqrt{2}}$, where $\ket{0}$ and $\ket{1}$ correspond to the top and bottom photon paths, respectively.
In this study, we replace the original control state $\ket{+}$ with a Bell state, ensuring that both photons are either in the top or bottom layer.

In order to realize the role of Fredkin gates, we design a new optical component that transmits the top-layer photon and reflects the bottom-layer photon, as shown in the inset of Fig.~\ref{fig:setup}(c).
We name this optical component as the spatial beam splitter (SBS), as it works like a polarization beam splitter (PBS) in spatial DoF.
In our experiment, when both photons are in the top layer, they transmit through the SBS, that is, $\rho_{\mathrm{m}}$ travels through $\mathcal{C}_1$ and $\rho$ travels through $\mathcal{C}_2$. 
When both photons are in the bottom layer, the SBS exchanges their trajectories, with $\rho_{\mathrm{m}}$ traveling through $\mathcal{C}_2$ and $\rho$ traveling through $\mathcal{C}_1$.
In this way, the SBS can exchange two target states according to the control state $\ket{\Phi^+}$, achieving the function of the Fredkin gate.
In fact, the SBS also exchanges the two control qubits according to their spatial information, as shown in the light blue box in Fig.~\ref{fig:setup}(b).
Nonetheless, as the two spatial qubits are confined to the subspace spanned by $\ket{00}$ and $\ket{11}$, they are not affected by the additional exchange operation.
After the two noise channels, the two photons pass through the SBS again to implement the second Fredkin gate.
At the end of the circuit, two qubits in spatial DoF are measured in Pauli-$X$ basis to post-select the measurement results.

Figure~\ref{fig:setup}(c) illustrates the detailed experimental setup.
In the initialization part, the entangled state $\ket{\Phi^+}$ is generated using an interference-based beam-like spontaneous parametric down conversion entanglement source~\cite{PhysRevLett.121.250505} in spatial DoF.
The maximally mixed state $\rho_{\mathrm{m}}$ in the ancillary register is prepared by dephasing the state $\ket{\text{D}}=\left(\ket{\text{H}}+\ket{\text{V}}\right)/{\sqrt{2}}$ with a thick yttrium vanadate (YVO$_4$) crystal, where $\ket{\text{H}}$ and $\ket{\text{V}}$ represent the horizontal and vertical polarization states of photons, respectively.
In the channel purification part, two noise channels $\mathcal{C}_1$ and $\mathcal{C}_2$ are simulated with the wave plate combination of two quarter-wave plates (QWPs) and a half-wave plate (HWP), which can be used to implement single-qubit Pauli rotations.
Four Pauli rotations are independently applied with certain probabilities to simulate noise channels.
In the measurement part, the beam splitter (BS) acts as the Hadamard gate in spatial DoF and is used to execute Pauli-$X$ measurement.
The purification efficiency of this protocol depends on the interference visibility in spatial DoF.
In the experiment, the visibility reaches 0.936 when two Pauli rotations are identical, which ensures the efficient implementation of our protocol.

\textbf{Results.}
To evaluate the performance of the channel purification protocol, we perform process tomography measurement on the two initial noise channels and the purified channels. 
Since process tomography measurement requires a total of 12 state preparation and measurement settings, we use stepping motors for rapid adjustments of all HWPs and QWPs. 
In each trial, we record the measurement results of the two spatial qubits and the polarization qubit in the main register. 
By post-selecting and post-processing the spatial DoF measurements, we extract the physically purified channel $\mathcal{C}_\mathrm{p}$ and the virtually purified channel $\mathcal{C}_{\mathrm{vp}}$.

\begin{figure}[htbp!]
	\centering
    \includegraphics[width=1 \linewidth]{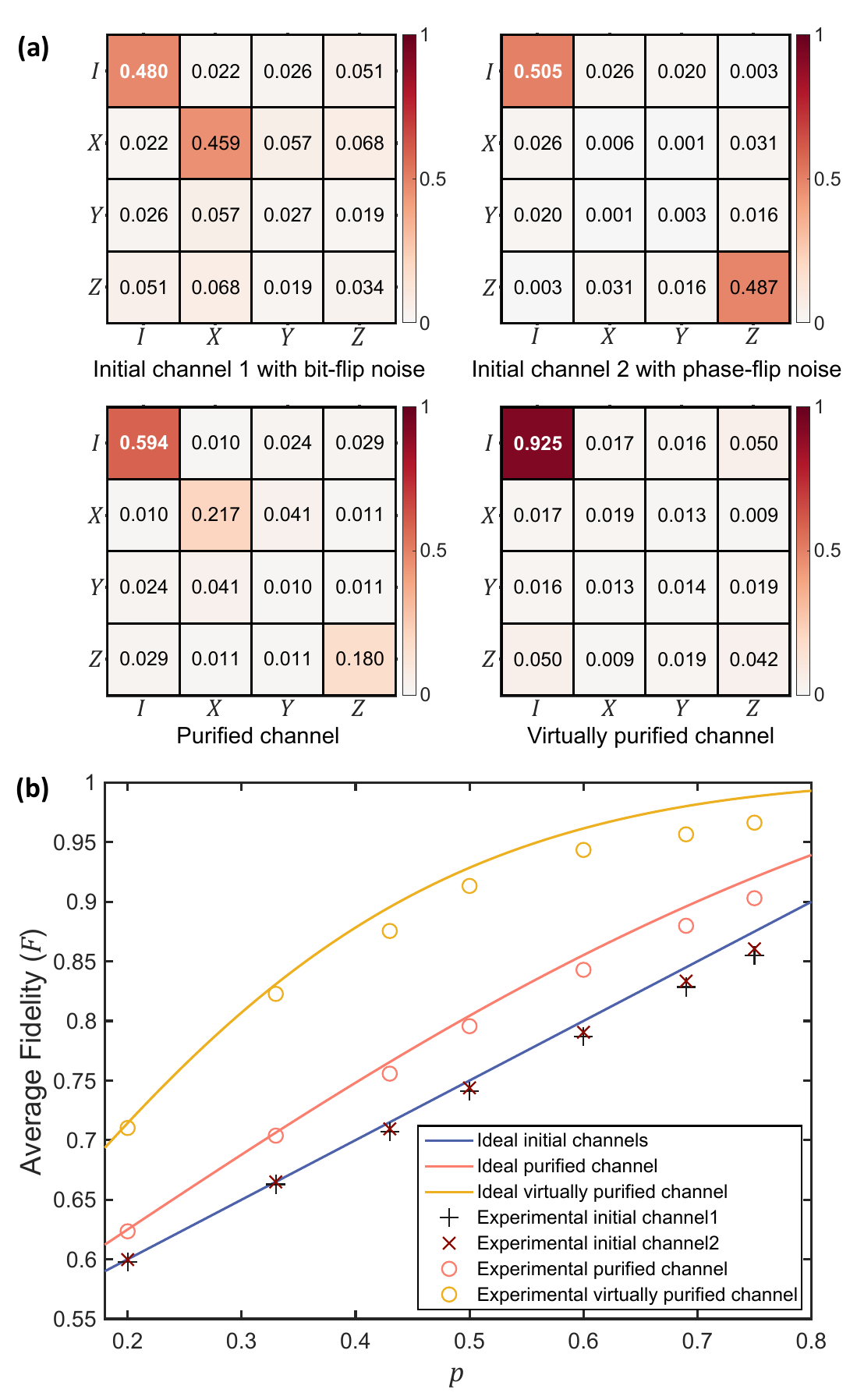}
\caption{Experimental results for channel purification.
(a) Pauli transfer matrices (magnitude) of the initial channels and the purified channels.
(b) Average state fidelities between initial and purified channels with the ideal noiseless channel. 
Error bar here is about $10^{-4}$. 
%Deviation between experimental and theoretical results of virtual purification for large $p$ is mainly due to the collection efficiency of detectors are unbalanced.(Delete)
}
\label{fig:result1}
\end{figure}

In our experiment, we first choose two different noise channels, where $\mathcal{C}_1$ and $\mathcal{C}_2$ are identity channels mixed with bit-flip and phase-flip errors, respectively.
The results of maximum likelihood process tomography measurement on these two noise channels~\cite{PhysRevA.63.020101} are shown in Fig.~\ref{fig:result1}(a). 
The matrix element $(I,I)$ represents the noiseless component of the channel.
After purification, this value increases from $0.480$ and $0.505$ to $0.594$.
With virtual purification, the element $(I,I)$ is further enhanced to $0.925$.
Note that the off-diagonal terms of the Pauli transfer matrix show no significant change after purification.
This observation is in line with our theoretical conclusion: the protocol preserves the Pauli-diagonal structure and does not introduce spurious noise components.

To generalize these findings, we calculate the average fidelity~\cite{nielsen2002simple} with the noiseless channel by setting two initial noise channels to be the depolarizing channels,
\begin{equation}\label{eq:noisy_channel}
\mathcal{C}(\rho)=p\rho+(1-p)\rho_{\mathrm{m}},
\end{equation} 
which is a typical noise channel in quantum networks that simultaneously introduces multiple error components.
In our demonstration, the experimental results closely match the theoretical curves shown in Fig.~\ref{fig:result1}(b), benefiting from the stability of the experimental setup.
Both two purification methods achieve a higher average fidelity over a wide range, from $p=0.2$ to $0.75$, with a maximal improvement from $0.744$ to $0.913$ by virtual purification.
Combined with the data in Fig.~\ref{fig:result1}(a), these results show the wide applicability and the strong noise suppression capability of our experimental architecture.

\begin{figure}[htbp!]
	\centering
\includegraphics[width=1 \linewidth]{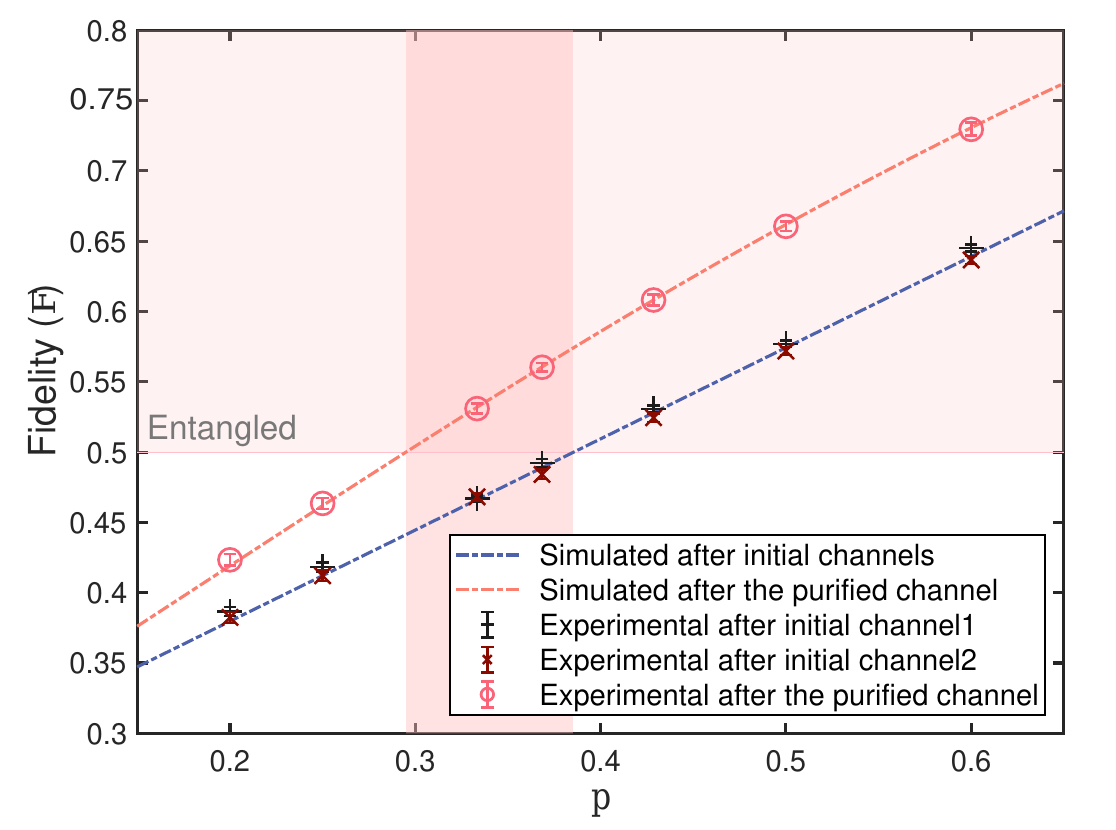}
\caption{Experimental results for entanglement distribution.
Fidelity is quantified as the overlap between the distributed state and the Bell state $\ket{\Phi^+}$.
The simulated lines are based on the initial state tomographic data without introducing noise. 
The fidelity larger than $0.5$ indicates the existence of entanglement.
In the range of $p$ labeled by the red rectangle, the entanglement shared by two parties is destroyed by the noise channel, while it can be preserved by channel purification.
}
\label{fig:result2}
\end{figure}

We further demonstrate the practical application of our channel purification protocol in entanglement distribution, which is a key task in quantum networks~\cite{doi:10.1126/science.aan3211}. 
Similarly to Fig.~\ref{fig:protocol}(b), we consider a bipartite entanglement distribution task in which a quantum network node sends a Bell state to two clients via noise channels.
For simplicity, we restrict our analysis to the case where the noise is mainly introduced by the channel between sender and Bob, specifically the depolarizing channel as defined in Eq.~\eqref{eq:noisy_channel}.
The experimental details are left in the Supplementary Material~\cite{SM}.

Finally, we demonstrate the performance of channel purification with the fidelity between the distributed state and the target Bell state $F=\bra{\Phi^+}\rho\ket{\Phi^+}$~\cite{PhysRevA.76.030305}.
As shown in Fig.~\ref{fig:result2}, for initial channels with a wide range of noise levels, the purified channel can always transmit states with the higher state fidelity.
In particular, for initial noise channels with the parameter $p$ in the red shaded range where transmitted entangled states degrade to separable states, the purified channel can preserve quantum entanglement.
For instance, for the initial noise channels with $p=0.33$, the distributed Bell states are verified as separable via the PPT criterion~\cite{peres1996ppt} after the transmission.
However, by applying purification operations to initial noise channels, the distributed state retains the entanglement with a fidelity of $0.528(3)$ after transmitting the purified channel.
This phenomenon highlights one inherent advantage of channel purification protocols over entanglement purification protocols, as the latter fundamentally cannot convert separable states into an entangled state \cite{bennett1996purification,Pan2003purification}.

\textbf{Discussion.}
In this study, we report the first experimental implementation of a channel purification protocol and its successful application in entanglement distribution. 
Our setup employs a spatial measurement configuration that operates as a Hong-Ou-Mandel interferometer~\cite{PhysRevA.87.052330}, which mitigates global phase fluctuations and suggests its potential for long-distance communication.
Future research will focus on adapting this architecture for more complex, multi-channel and multi-level purification scenarios, as well as exploring alternative protocols for enhanced performance and reduced implementation costs.

The channel purification protocol we use is similar to the protocol based on channel superposition~\cite{chiribella2019quantum,Abbott2020communication,rubino2021trajectories}, which replaces the maximally mixed state of Fig.~\ref{fig:setup}(a) with the vacuum state.
In contrast, the performance of our protocol depends solely on the noise channel itself, whereas that of protocols based on channel superposition also depends on the implementation of the noise channel, potentially conflicting with the requirements of channel purification.
Nevertheless, the relationship between these two protocols provides a new insight that helps us understand the physical essence of channel purification and to design new protocols.

\textbf{Acknowledgements.}
This work was supported by the National Natural Science Foundation of China (Grants No.~11975222, No.~11874340, No.~12174216), Shanghai Municipal Science and Technology Major Project (Grant No.~2019SHZDZX01), Chinese Academy of Sciences and the Shanghai Science and Technology Development Funds (Grant No.~18JC1414700), the Innovation Program for Quantum Science and Technology (Grant No.~2021ZD0301901, No.~2021ZD0300804), the Basic Science Center Project of NSFC (Grant No. 12488301), and the New Cornerstone Science Foundation.
Zhenyu Cai acknowledges support from the Junior Research Fellowship at St John’s College, Oxford.
Xu-Fei Yin was support from the China Postdoctoral Science Foundation (Grant No.~2023M733418).


%\bibliography{Ref_CP}

\begin{thebibliography}{41}%
	\makeatletter
	\providecommand \@ifxundefined [1]{%
		\@ifx{#1\undefined}
	}%
	\providecommand \@ifnum [1]{%
		\ifnum #1\expandafter \@firstoftwo
		\else \expandafter \@secondoftwo
		\fi
	}%
	\providecommand \@ifx [1]{%
		\ifx #1\expandafter \@firstoftwo
		\else \expandafter \@secondoftwo
		\fi
	}%
	\providecommand \natexlab [1]{#1}%
	\providecommand \enquote  [1]{``#1''}%
	\providecommand \bibnamefont  [1]{#1}%
	\providecommand \bibfnamefont [1]{#1}%
	\providecommand \citenamefont [1]{#1}%
	\providecommand \href@noop [0]{\@secondoftwo}%
	\providecommand \href [0]{\begingroup \@sanitize@url \@href}%
	\providecommand \@href[1]{\@@startlink{#1}\@@href}%
	\providecommand \@@href[1]{\endgroup#1\@@endlink}%
	\providecommand \@sanitize@url [0]{\catcode `\\12\catcode `\$12\catcode
		`\&12\catcode `\#12\catcode `\^12\catcode `\_12\catcode `\%12\relax}%
	\providecommand \@@startlink[1]{}%
	\providecommand \@@endlink[0]{}%
	\providecommand \url  [0]{\begingroup\@sanitize@url \@url }%
	\providecommand \@url [1]{\endgroup\@href {#1}{\urlprefix }}%
	\providecommand \urlprefix  [0]{URL }%
	\providecommand \Eprint [0]{\href }%
	\providecommand \doibase [0]{https://doi.org/}%
	\providecommand \selectlanguage [0]{\@gobble}%
	\providecommand \bibinfo  [0]{\@secondoftwo}%
	\providecommand \bibfield  [0]{\@secondoftwo}%
	\providecommand \translation [1]{[#1]}%
	\providecommand \BibitemOpen [0]{}%
	\providecommand \bibitemStop [0]{}%
	\providecommand \bibitemNoStop [0]{.\EOS\space}%
	\providecommand \EOS [0]{\spacefactor3000\relax}%
	\providecommand \BibitemShut  [1]{\csname bibitem#1\endcsname}%
	\let\auto@bib@innerbib\@empty
	%</preamble>
	\bibitem [{\citenamefont {Simon}(2017)}]{Simon2017network}%
	\BibitemOpen
	\bibfield  {author} {\bibinfo {author} {\bibfnamefont {C.}~\bibnamefont
			{Simon}},\ }\bibfield  {title} {\bibinfo {title} {Towards a global quantum
			network},\ }\href {https://www.nature.com/articles/s41566-017-0032-0}
	{\bibfield  {journal} {\bibinfo  {journal} {Nat. Photonics}\ }\textbf
		{\bibinfo {volume} {11}},\ \bibinfo {pages} {678} (\bibinfo {year}
		{2017})}\BibitemShut {NoStop}%
	\bibitem [{\citenamefont {Chen}\ \emph {et~al.}(2021)\citenamefont {Chen},
		\citenamefont {Zhang}, \citenamefont {Chen}, \citenamefont {Cai},
		\citenamefont {Liao}, \citenamefont {Zhang}, \citenamefont {Chen},
		\citenamefont {Yin}, \citenamefont {Ren}, \citenamefont {Chen} \emph
		{et~al.}}]{chen2021integrated}%
	\BibitemOpen
	\bibfield  {author} {\bibinfo {author} {\bibfnamefont {Y.-A.}\ \bibnamefont
			{Chen}}, \bibinfo {author} {\bibfnamefont {Q.}~\bibnamefont {Zhang}},
		\bibinfo {author} {\bibfnamefont {T.-Y.}\ \bibnamefont {Chen}}, \bibinfo
		{author} {\bibfnamefont {W.-Q.}\ \bibnamefont {Cai}}, \bibinfo {author}
		{\bibfnamefont {S.-K.}\ \bibnamefont {Liao}}, \bibinfo {author}
		{\bibfnamefont {J.}~\bibnamefont {Zhang}}, \bibinfo {author} {\bibfnamefont
			{K.}~\bibnamefont {Chen}}, \bibinfo {author} {\bibfnamefont {J.}~\bibnamefont
			{Yin}}, \bibinfo {author} {\bibfnamefont {J.-G.}\ \bibnamefont {Ren}},
		\bibinfo {author} {\bibfnamefont {Z.}~\bibnamefont {Chen}}, \emph {et~al.},\
	}\bibfield  {title} {\bibinfo {title} {An integrated space-to-ground quantum
			communication network over 4,600 kilometres},\ }\href
	{https://doi.org/10.1038/s41586-020-03093-8} {\bibfield  {journal} {\bibinfo
			{journal} {Nature}\ }\textbf {\bibinfo {volume} {589}},\ \bibinfo {pages}
		{214} (\bibinfo {year} {2021})}\BibitemShut {NoStop}%
	\bibitem [{\citenamefont {Main}\ \emph {et~al.}(2025)\citenamefont {Main},
		\citenamefont {Drmota}, \citenamefont {Nadlinger}, \citenamefont {Ainley},
		\citenamefont {Agrawal}, \citenamefont {Nichol}, \citenamefont {Srinivas},
		\citenamefont {Araneda},\ and\ \citenamefont {Lucas}}]{main2025distributed}%
	\BibitemOpen
	\bibfield  {author} {\bibinfo {author} {\bibfnamefont {D.}~\bibnamefont
			{Main}}, \bibinfo {author} {\bibfnamefont {P.}~\bibnamefont {Drmota}},
		\bibinfo {author} {\bibfnamefont {D.}~\bibnamefont {Nadlinger}}, \bibinfo
		{author} {\bibfnamefont {E.}~\bibnamefont {Ainley}}, \bibinfo {author}
		{\bibfnamefont {A.}~\bibnamefont {Agrawal}}, \bibinfo {author} {\bibfnamefont
			{B.}~\bibnamefont {Nichol}}, \bibinfo {author} {\bibfnamefont
			{R.}~\bibnamefont {Srinivas}}, \bibinfo {author} {\bibfnamefont
			{G.}~\bibnamefont {Araneda}},\ and\ \bibinfo {author} {\bibfnamefont
			{D.}~\bibnamefont {Lucas}},\ }\bibfield  {title} {\bibinfo {title}
		{Distributed quantum computing across an optical network link},\ }\href
	{https://doi.org/10.1038/s41586-024-08404-x} {\bibfield  {journal} {\bibinfo
			{journal} {Nature}\ ,\ \bibinfo {pages} {1}} (\bibinfo {year}
		{2025})}\BibitemShut {NoStop}%
	\bibitem [{\citenamefont {Pan}\ \emph {et~al.}(2012)\citenamefont {Pan},
		\citenamefont {Chen}, \citenamefont {Lu}, \citenamefont {Weinfurter},
		\citenamefont {Zeilinger},\ and\ \citenamefont {\ifmmode~\dot{Z}\else
			\.{Z}\fi{}ukowski}}]{pan2012multiphoton}%
	\BibitemOpen
	\bibfield  {author} {\bibinfo {author} {\bibfnamefont {J.-W.}\ \bibnamefont
			{Pan}}, \bibinfo {author} {\bibfnamefont {Z.-B.}\ \bibnamefont {Chen}},
		\bibinfo {author} {\bibfnamefont {C.-Y.}\ \bibnamefont {Lu}}, \bibinfo
		{author} {\bibfnamefont {H.}~\bibnamefont {Weinfurter}}, \bibinfo {author}
		{\bibfnamefont {A.}~\bibnamefont {Zeilinger}},\ and\ \bibinfo {author}
		{\bibfnamefont {M.}~\bibnamefont {\ifmmode~\dot{Z}\else \.{Z}\fi{}ukowski}},\
	}\bibfield  {title} {\bibinfo {title} {Multiphoton entanglement and
			interferometry},\ }\href
	{https://journals.aps.org/rmp/abstract/10.1103/RevModPhys.84.777} {\bibfield
		{journal} {\bibinfo  {journal} {Rev. Mod. Phys.}\ }\textbf {\bibinfo {volume}
			{84}},\ \bibinfo {pages} {777} (\bibinfo {year} {2012})}\BibitemShut
	{NoStop}%
	\bibitem [{\citenamefont {Devitt}\ \emph {et~al.}(2013)\citenamefont {Devitt},
		\citenamefont {Munro},\ and\ \citenamefont {Nemoto}}]{Devitt2013qec}%
	\BibitemOpen
	\bibfield  {author} {\bibinfo {author} {\bibfnamefont {S.~J.}\ \bibnamefont
			{Devitt}}, \bibinfo {author} {\bibfnamefont {W.~J.}\ \bibnamefont {Munro}},\
		and\ \bibinfo {author} {\bibfnamefont {K.}~\bibnamefont {Nemoto}},\
	}\bibfield  {title} {\bibinfo {title} {Quantum error correction for
			beginners},\ }\href {https://doi.org/10.1088/0034-4885/76/7/076001}
	{\bibfield  {journal} {\bibinfo  {journal} {Reports on Progress in Physics}\
		}\textbf {\bibinfo {volume} {76}},\ \bibinfo {pages} {076001} (\bibinfo
		{year} {2013})}\BibitemShut {NoStop}%
	\bibitem [{\citenamefont {Terhal}(2015)}]{terhal2015memory}%
	\BibitemOpen
	\bibfield  {author} {\bibinfo {author} {\bibfnamefont {B.~M.}\ \bibnamefont
			{Terhal}},\ }\bibfield  {title} {\bibinfo {title} {Quantum error correction
			for quantum memories},\ }\href {https://doi.org/10.1103/RevModPhys.87.307}
	{\bibfield  {journal} {\bibinfo  {journal} {Rev. Mod. Phys.}\ }\textbf
		{\bibinfo {volume} {87}},\ \bibinfo {pages} {307} (\bibinfo {year}
		{2015})}\BibitemShut {NoStop}%
	\bibitem [{\citenamefont {Bennett}\ \emph {et~al.}(1996)\citenamefont
		{Bennett}, \citenamefont {Brassard}, \citenamefont {Popescu}, \citenamefont
		{Schumacher}, \citenamefont {Smolin},\ and\ \citenamefont
		{Wootters}}]{bennett1996purification}%
	\BibitemOpen
	\bibfield  {author} {\bibinfo {author} {\bibfnamefont {C.~H.}\ \bibnamefont
			{Bennett}}, \bibinfo {author} {\bibfnamefont {G.}~\bibnamefont {Brassard}},
		\bibinfo {author} {\bibfnamefont {S.}~\bibnamefont {Popescu}}, \bibinfo
		{author} {\bibfnamefont {B.}~\bibnamefont {Schumacher}}, \bibinfo {author}
		{\bibfnamefont {J.~A.}\ \bibnamefont {Smolin}},\ and\ \bibinfo {author}
		{\bibfnamefont {W.~K.}\ \bibnamefont {Wootters}},\ }\bibfield  {title}
	{\bibinfo {title} {Purification of noisy entanglement and faithful
			teleportation via noisy channels},\ }\href
	{https://doi.org/10.1103/PhysRevLett.76.722} {\bibfield  {journal} {\bibinfo
			{journal} {Phys. Rev. Lett.}\ }\textbf {\bibinfo {volume} {76}},\ \bibinfo
		{pages} {722} (\bibinfo {year} {1996})}\BibitemShut {NoStop}%
	\bibitem [{\citenamefont {Pan}\ \emph {et~al.}(2001)\citenamefont {Pan},
		\citenamefont {Simon}, \citenamefont {Brukner},\ and\ \citenamefont
		{Zeilinger}}]{Pan2001purification}%
	\BibitemOpen
	\bibfield  {author} {\bibinfo {author} {\bibfnamefont {J.-W.}\ \bibnamefont
			{Pan}}, \bibinfo {author} {\bibfnamefont {C.}~\bibnamefont {Simon}}, \bibinfo
		{author} {\bibfnamefont {{\v{C}}.}~\bibnamefont {Brukner}},\ and\ \bibinfo
		{author} {\bibfnamefont {A.}~\bibnamefont {Zeilinger}},\ }\bibfield  {title}
	{\bibinfo {title} {Entanglement purification for quantum communication},\
	}\href {https://www.nature.com/articles/35074041} {\bibfield  {journal}
		{\bibinfo  {journal} {Nature}\ }\textbf {\bibinfo {volume} {410}},\ \bibinfo
		{pages} {1067} (\bibinfo {year} {2001})}\BibitemShut {NoStop}%
	\bibitem [{\citenamefont {Ni}\ \emph {et~al.}(2023)\citenamefont {Ni},
		\citenamefont {Li}, \citenamefont {Deng}, \citenamefont {Cai}, \citenamefont
		{Zhang}, \citenamefont {Wang}, \citenamefont {Yang}, \citenamefont {Yu},
		\citenamefont {Yan}, \citenamefont {Liu}, \citenamefont {Zou}, \citenamefont
		{Sun}, \citenamefont {Zheng}, \citenamefont {Xu},\ and\ \citenamefont
		{Yu}}]{Ni2023breakeven}%
	\BibitemOpen
	\bibfield  {author} {\bibinfo {author} {\bibfnamefont {Z.}~\bibnamefont
			{Ni}}, \bibinfo {author} {\bibfnamefont {S.}~\bibnamefont {Li}}, \bibinfo
		{author} {\bibfnamefont {X.}~\bibnamefont {Deng}}, \bibinfo {author}
		{\bibfnamefont {Y.}~\bibnamefont {Cai}}, \bibinfo {author} {\bibfnamefont
			{L.}~\bibnamefont {Zhang}}, \bibinfo {author} {\bibfnamefont
			{W.}~\bibnamefont {Wang}}, \bibinfo {author} {\bibfnamefont {Z.-B.}\
			\bibnamefont {Yang}}, \bibinfo {author} {\bibfnamefont {H.}~\bibnamefont
			{Yu}}, \bibinfo {author} {\bibfnamefont {F.}~\bibnamefont {Yan}}, \bibinfo
		{author} {\bibfnamefont {S.}~\bibnamefont {Liu}}, \bibinfo {author}
		{\bibfnamefont {C.-L.}\ \bibnamefont {Zou}}, \bibinfo {author} {\bibfnamefont
			{L.}~\bibnamefont {Sun}}, \bibinfo {author} {\bibfnamefont {S.-B.}\
			\bibnamefont {Zheng}}, \bibinfo {author} {\bibfnamefont {Y.}~\bibnamefont
			{Xu}},\ and\ \bibinfo {author} {\bibfnamefont {D.}~\bibnamefont {Yu}},\
	}\bibfield  {title} {\bibinfo {title} {Beating the break-even point with a
			discrete-variable-encoded logical qubit},\ }\href
	{https://doi.org/10.1038/s41586-023-05784-4} {\bibfield  {journal} {\bibinfo
			{journal} {Nature}\ }\textbf {\bibinfo {volume} {616}},\ \bibinfo {pages}
		{56} (\bibinfo {year} {2023})}\BibitemShut {NoStop}%
	\bibitem [{\citenamefont {Bluvstein}\ \emph {et~al.}(2024)\citenamefont
		{Bluvstein}, \citenamefont {Evered}, \citenamefont {Geim}, \citenamefont
		{Li}, \citenamefont {Zhou}, \citenamefont {Manovitz}, \citenamefont {Ebadi},
		\citenamefont {Cain}, \citenamefont {Kalinowski}, \citenamefont {Hangleiter},
		\citenamefont {Bonilla~Ataides}, \citenamefont {Maskara}, \citenamefont
		{Cong}, \citenamefont {Gao}, \citenamefont {Sales~Rodriguez}, \citenamefont
		{Karolyshyn}, \citenamefont {Semeghini}, \citenamefont {Gullans},
		\citenamefont {Greiner}, \citenamefont {Vuletić},\ and\ \citenamefont
		{Lukin}}]{Bluvstein2024qec}%
	\BibitemOpen
	\bibfield  {author} {\bibinfo {author} {\bibfnamefont {D.}~\bibnamefont
			{Bluvstein}}, \bibinfo {author} {\bibfnamefont {S.~J.}\ \bibnamefont
			{Evered}}, \bibinfo {author} {\bibfnamefont {A.~A.}\ \bibnamefont {Geim}},
		\bibinfo {author} {\bibfnamefont {S.~H.}\ \bibnamefont {Li}}, \bibinfo
		{author} {\bibfnamefont {H.}~\bibnamefont {Zhou}}, \bibinfo {author}
		{\bibfnamefont {T.}~\bibnamefont {Manovitz}}, \bibinfo {author}
		{\bibfnamefont {S.}~\bibnamefont {Ebadi}}, \bibinfo {author} {\bibfnamefont
			{M.}~\bibnamefont {Cain}}, \bibinfo {author} {\bibfnamefont {M.}~\bibnamefont
			{Kalinowski}}, \bibinfo {author} {\bibfnamefont {D.}~\bibnamefont
			{Hangleiter}}, \bibinfo {author} {\bibfnamefont {J.~P.}\ \bibnamefont
			{Bonilla~Ataides}}, \bibinfo {author} {\bibfnamefont {N.}~\bibnamefont
			{Maskara}}, \bibinfo {author} {\bibfnamefont {I.}~\bibnamefont {Cong}},
		\bibinfo {author} {\bibfnamefont {X.}~\bibnamefont {Gao}}, \bibinfo {author}
		{\bibfnamefont {P.}~\bibnamefont {Sales~Rodriguez}}, \bibinfo {author}
		{\bibfnamefont {T.}~\bibnamefont {Karolyshyn}}, \bibinfo {author}
		{\bibfnamefont {G.}~\bibnamefont {Semeghini}}, \bibinfo {author}
		{\bibfnamefont {M.~J.}\ \bibnamefont {Gullans}}, \bibinfo {author}
		{\bibfnamefont {M.}~\bibnamefont {Greiner}}, \bibinfo {author} {\bibfnamefont
			{V.}~\bibnamefont {Vuletić}},\ and\ \bibinfo {author} {\bibfnamefont
			{M.~D.}\ \bibnamefont {Lukin}},\ }\bibfield  {title} {\bibinfo {title}
		{Logical quantum processor based on reconfigurable atom arrays},\ }\href
	{https://doi.org/10.1038/s41586-023-06927-3} {\bibfield  {journal} {\bibinfo
			{journal} {Nature}\ }\textbf {\bibinfo {volume} {626}},\ \bibinfo {pages}
		{58} (\bibinfo {year} {2024})}\BibitemShut {NoStop}%
	\bibitem [{\citenamefont {Acharya}\ \emph {et~al.}(2025)\citenamefont
		{Acharya}, \citenamefont {Aghababaie-Beni}, \citenamefont {Aleiner},
		\citenamefont {Andersen}, \citenamefont {Ansmann}, \citenamefont {Arute},
		\citenamefont {Arya}, \citenamefont {Asfaw}, \citenamefont {Astrakhantsev},
		\citenamefont {Atalaya} \emph {et~al.}}]{Acharya2025surface}%
	\BibitemOpen
	\bibfield  {author} {\bibinfo {author} {\bibfnamefont {R.}~\bibnamefont
			{Acharya}}, \bibinfo {author} {\bibfnamefont {L.}~\bibnamefont
			{Aghababaie-Beni}}, \bibinfo {author} {\bibfnamefont {I.}~\bibnamefont
			{Aleiner}}, \bibinfo {author} {\bibfnamefont {T.~I.}\ \bibnamefont
			{Andersen}}, \bibinfo {author} {\bibfnamefont {M.}~\bibnamefont {Ansmann}},
		\bibinfo {author} {\bibfnamefont {F.}~\bibnamefont {Arute}}, \bibinfo
		{author} {\bibfnamefont {K.}~\bibnamefont {Arya}}, \bibinfo {author}
		{\bibfnamefont {A.}~\bibnamefont {Asfaw}}, \bibinfo {author} {\bibfnamefont
			{N.}~\bibnamefont {Astrakhantsev}}, \bibinfo {author} {\bibfnamefont
			{J.}~\bibnamefont {Atalaya}}, \emph {et~al.},\ }\bibfield  {title} {\bibinfo
		{title} {Quantum error correction below the surface code threshold},\ }\href
	{https://doi.org/10.1038/s41586-024-08449-y} {\bibfield  {journal} {\bibinfo
			{journal} {Nature}\ }\textbf {\bibinfo {volume} {638}},\ \bibinfo {pages}
		{920} (\bibinfo {year} {2025})}\BibitemShut {NoStop}%
	\bibitem [{\citenamefont {Liu}\ \emph {et~al.}(2025)\citenamefont {Liu},
		\citenamefont {Zhang}, \citenamefont {Fei},\ and\ \citenamefont
		{Cai}}]{liu2024virtualchannelpurification}%
	\BibitemOpen
	\bibfield  {author} {\bibinfo {author} {\bibfnamefont {Z.}~\bibnamefont
			{Liu}}, \bibinfo {author} {\bibfnamefont {X.}~\bibnamefont {Zhang}}, \bibinfo
		{author} {\bibfnamefont {Y.-Y.}\ \bibnamefont {Fei}},\ and\ \bibinfo {author}
		{\bibfnamefont {Z.}~\bibnamefont {Cai}},\ }\bibfield  {title} {\bibinfo
		{title} {Virtual channel purification},\ }\href
	{https://doi.org/10.1103/PRXQuantum.6.020325} {\bibfield  {journal} {\bibinfo
			{journal} {PRX Quantum}\ }\textbf {\bibinfo {volume} {6}},\ \bibinfo {pages}
		{020325} (\bibinfo {year} {2025})}\BibitemShut {NoStop}%
	\bibitem [{\citenamefont {Miguel-Ramiro}\ \emph {et~al.}(2023)\citenamefont
		{Miguel-Ramiro}, \citenamefont {Shi}, \citenamefont {Dellantonio},
		\citenamefont {Chan}, \citenamefont {Muschik},\ and\ \citenamefont
		{D\"ur}}]{PhysRevLett.131.230601}%
	\BibitemOpen
	\bibfield  {author} {\bibinfo {author} {\bibfnamefont {J.}~\bibnamefont
			{Miguel-Ramiro}}, \bibinfo {author} {\bibfnamefont {Z.}~\bibnamefont {Shi}},
		\bibinfo {author} {\bibfnamefont {L.}~\bibnamefont {Dellantonio}}, \bibinfo
		{author} {\bibfnamefont {A.}~\bibnamefont {Chan}}, \bibinfo {author}
		{\bibfnamefont {C.~A.}\ \bibnamefont {Muschik}},\ and\ \bibinfo {author}
		{\bibfnamefont {W.}~\bibnamefont {D\"ur}},\ }\bibfield  {title} {\bibinfo
		{title} {Superposed quantum error mitigation},\ }\href
	{https://doi.org/10.1103/PhysRevLett.131.230601} {\bibfield  {journal}
		{\bibinfo  {journal} {Phys. Rev. Lett.}\ }\textbf {\bibinfo {volume} {131}},\
		\bibinfo {pages} {230601} (\bibinfo {year} {2023})}\BibitemShut {NoStop}%
	\bibitem [{\citenamefont {Huang}\ and\ \citenamefont
		{Lupo}(2024)}]{PhysRevA.110.052431}%
	\BibitemOpen
	\bibfield  {author} {\bibinfo {author} {\bibfnamefont {Z.}~\bibnamefont
			{Huang}}\ and\ \bibinfo {author} {\bibfnamefont {C.}~\bibnamefont {Lupo}},\
	}\bibfield  {title} {\bibinfo {title} {Error filtration for quantum sensing
			via interferometry},\ }\href {https://doi.org/10.1103/PhysRevA.110.052431}
	{\bibfield  {journal} {\bibinfo  {journal} {Phys. Rev. A}\ }\textbf {\bibinfo
			{volume} {110}},\ \bibinfo {pages} {052431} (\bibinfo {year}
		{2024})}\BibitemShut {NoStop}%
	\bibitem [{\citenamefont {Cirac}\ \emph {et~al.}(1999)\citenamefont {Cirac},
		\citenamefont {Ekert},\ and\ \citenamefont
		{Macchiavello}}]{cirac1999optimal}%
	\BibitemOpen
	\bibfield  {author} {\bibinfo {author} {\bibfnamefont {J.~I.}\ \bibnamefont
			{Cirac}}, \bibinfo {author} {\bibfnamefont {A.~K.}\ \bibnamefont {Ekert}},\
		and\ \bibinfo {author} {\bibfnamefont {C.}~\bibnamefont {Macchiavello}},\
	}\bibfield  {title} {\bibinfo {title} {Optimal purification of single
			qubits},\ }\href {https://doi.org/10.1103/PhysRevLett.82.4344} {\bibfield
		{journal} {\bibinfo  {journal} {Phys. Rev. Lett.}\ }\textbf {\bibinfo
			{volume} {82}},\ \bibinfo {pages} {4344} (\bibinfo {year}
		{1999})}\BibitemShut {NoStop}%
	\bibitem [{\citenamefont {Cotler}\ \emph {et~al.}(2019)\citenamefont {Cotler},
		\citenamefont {Choi}, \citenamefont {Lukin}, \citenamefont {Gharibyan},
		\citenamefont {Grover}, \citenamefont {Tai}, \citenamefont {Rispoli},
		\citenamefont {Schittko}, \citenamefont {Preiss}, \citenamefont {Kaufman},
		\citenamefont {Greiner}, \citenamefont {Pichler},\ and\ \citenamefont
		{Hayden}}]{cotler2019cooling}%
	\BibitemOpen
	\bibfield  {author} {\bibinfo {author} {\bibfnamefont {J.}~\bibnamefont
			{Cotler}}, \bibinfo {author} {\bibfnamefont {S.}~\bibnamefont {Choi}},
		\bibinfo {author} {\bibfnamefont {A.}~\bibnamefont {Lukin}}, \bibinfo
		{author} {\bibfnamefont {H.}~\bibnamefont {Gharibyan}}, \bibinfo {author}
		{\bibfnamefont {T.}~\bibnamefont {Grover}}, \bibinfo {author} {\bibfnamefont
			{M.~E.}\ \bibnamefont {Tai}}, \bibinfo {author} {\bibfnamefont
			{M.}~\bibnamefont {Rispoli}}, \bibinfo {author} {\bibfnamefont
			{R.}~\bibnamefont {Schittko}}, \bibinfo {author} {\bibfnamefont {P.~M.}\
			\bibnamefont {Preiss}}, \bibinfo {author} {\bibfnamefont {A.~M.}\
			\bibnamefont {Kaufman}}, \bibinfo {author} {\bibfnamefont {M.}~\bibnamefont
			{Greiner}}, \bibinfo {author} {\bibfnamefont {H.}~\bibnamefont {Pichler}},\
		and\ \bibinfo {author} {\bibfnamefont {P.}~\bibnamefont {Hayden}},\
	}\bibfield  {title} {\bibinfo {title} {Quantum virtual cooling},\ }\href
	{https://doi.org/10.1103/PhysRevX.9.031013} {\bibfield  {journal} {\bibinfo
			{journal} {Phys. Rev. X}\ }\textbf {\bibinfo {volume} {9}},\ \bibinfo {pages}
		{031013} (\bibinfo {year} {2019})}\BibitemShut {NoStop}%
	\bibitem [{\citenamefont {Huggins}\ \emph {et~al.}(2021)\citenamefont
		{Huggins}, \citenamefont {McArdle}, \citenamefont {O'Brien}, \citenamefont
		{Lee}, \citenamefont {Rubin}, \citenamefont {Boixo}, \citenamefont {Whaley},
		\citenamefont {Babbush},\ and\ \citenamefont
		{McClean}}]{hugginsVirtualDistillationQuantum2021}%
	\BibitemOpen
	\bibfield  {author} {\bibinfo {author} {\bibfnamefont {W.~J.}\ \bibnamefont
			{Huggins}}, \bibinfo {author} {\bibfnamefont {S.}~\bibnamefont {McArdle}},
		\bibinfo {author} {\bibfnamefont {T.~E.}\ \bibnamefont {O'Brien}}, \bibinfo
		{author} {\bibfnamefont {J.}~\bibnamefont {Lee}}, \bibinfo {author}
		{\bibfnamefont {N.~C.}\ \bibnamefont {Rubin}}, \bibinfo {author}
		{\bibfnamefont {S.}~\bibnamefont {Boixo}}, \bibinfo {author} {\bibfnamefont
			{K.~B.}\ \bibnamefont {Whaley}}, \bibinfo {author} {\bibfnamefont
			{R.}~\bibnamefont {Babbush}},\ and\ \bibinfo {author} {\bibfnamefont {J.~R.}\
			\bibnamefont {McClean}},\ }\bibfield  {title} {\bibinfo {title} {Virtual
			distillation for quantum error mitigation},\ }\href
	{https://doi.org/10.1103/PhysRevX.11.041036} {\bibfield  {journal} {\bibinfo
			{journal} {Phys. Rev. X}\ }\textbf {\bibinfo {volume} {11}},\ \bibinfo
		{pages} {41036} (\bibinfo {year} {2021})}\BibitemShut {NoStop}%
	\bibitem [{\citenamefont
		{Koczor}(2021)}]{koczorExponentialErrorSuppression2021}%
	\BibitemOpen
	\bibfield  {author} {\bibinfo {author} {\bibfnamefont {B.}~\bibnamefont
			{Koczor}},\ }\bibfield  {title} {\bibinfo {title} {Exponential error
			suppression for near-term quantum devices},\ }\href
	{https://doi.org/10.1103/PhysRevX.11.031057} {\bibfield  {journal} {\bibinfo
			{journal} {Phys. Rev. X}\ }\textbf {\bibinfo {volume} {11}},\ \bibinfo
		{pages} {31057} (\bibinfo {year} {2021})}\BibitemShut {NoStop}%
	\bibitem [{\citenamefont {Chiribella}\ and\ \citenamefont
		{Kristj{\'a}nsson}(2019)}]{chiribella2019quantum}%
	\BibitemOpen
	\bibfield  {author} {\bibinfo {author} {\bibfnamefont {G.}~\bibnamefont
			{Chiribella}}\ and\ \bibinfo {author} {\bibfnamefont {H.}~\bibnamefont
			{Kristj{\'a}nsson}},\ }\bibfield  {title} {\bibinfo {title} {Quantum shannon
			theory with superpositions of trajectories},\ }\href
	{https://royalsocietypublishing.org/doi/full/10.1098/rspa.2018.0903}
	{\bibfield  {journal} {\bibinfo  {journal} {Proceedings of the Royal Society
				A}\ }\textbf {\bibinfo {volume} {475}},\ \bibinfo {pages} {20180903}
		(\bibinfo {year} {2019})}\BibitemShut {NoStop}%
	\bibitem [{\citenamefont {Lloyd}\ \emph {et~al.}(2014)\citenamefont {Lloyd},
		\citenamefont {Mohseni},\ and\ \citenamefont {Rebentrost}}]{Lloyd2014qpca}%
	\BibitemOpen
	\bibfield  {author} {\bibinfo {author} {\bibfnamefont {S.}~\bibnamefont
			{Lloyd}}, \bibinfo {author} {\bibfnamefont {M.}~\bibnamefont {Mohseni}},\
		and\ \bibinfo {author} {\bibfnamefont {P.}~\bibnamefont {Rebentrost}},\
	}\bibfield  {title} {\bibinfo {title} {Quantum principal component
			analysis},\ }\href {https://doi.org/10.1038/nphys3029} {\bibfield  {journal}
		{\bibinfo  {journal} {Nature Physics}\ }\textbf {\bibinfo {volume} {10}},\
		\bibinfo {pages} {631} (\bibinfo {year} {2014})}\BibitemShut {NoStop}%
	\bibitem [{\citenamefont {Zhou}\ and\ \citenamefont
		{Liu}(2024)}]{Zhou2024hybrid}%
	\BibitemOpen
	\bibfield  {author} {\bibinfo {author} {\bibfnamefont {Y.}~\bibnamefont
			{Zhou}}\ and\ \bibinfo {author} {\bibfnamefont {Z.}~\bibnamefont {Liu}},\
	}\bibfield  {title} {\bibinfo {title} {A hybrid framework for estimating
			nonlinear functions of quantum states},\ }\href
	{https://doi.org/10.1038/s41534-024-00846-5} {\bibfield  {journal} {\bibinfo
			{journal} {npj Quantum Information}\ }\textbf {\bibinfo {volume} {10}},\
		\bibinfo {pages} {62} (\bibinfo {year} {2024})}\BibitemShut {NoStop}%
	\bibitem [{\citenamefont {Faehrmann}\ \emph {et~al.}(2025)\citenamefont
		{Faehrmann}, \citenamefont {Eisert},\ and\ \citenamefont
		{Kueng}}]{faehrmann2025hadamard}%
	\BibitemOpen
	\bibfield  {author} {\bibinfo {author} {\bibfnamefont {P.~K.}\ \bibnamefont
			{Faehrmann}}, \bibinfo {author} {\bibfnamefont {J.}~\bibnamefont {Eisert}},\
		and\ \bibinfo {author} {\bibfnamefont {R.}~\bibnamefont {Kueng}},\ }\bibfield
	{title} {\bibinfo {title} {In the shadow of the hadamard test: Using the
			garbage state for good and further modifications},\ }\href
	{https://doi.org/10.1103/cqjw-kl8s} {\bibfield  {journal} {\bibinfo
			{journal} {Phys. Rev. Lett.}\ }\textbf {\bibinfo {volume} {135}},\ \bibinfo
		{pages} {150603} (\bibinfo {year} {2025})}\BibitemShut {NoStop}%
	\bibitem [{\citenamefont {Wallman}\ and\ \citenamefont
		{Emerson}(2016)}]{wallmanNoiseTailoringScalable2016}%
	\BibitemOpen
	\bibfield  {author} {\bibinfo {author} {\bibfnamefont {J.~J.}\ \bibnamefont
			{Wallman}}\ and\ \bibinfo {author} {\bibfnamefont {J.}~\bibnamefont
			{Emerson}},\ }\bibfield  {title} {\bibinfo {title} {Noise tailoring for
			scalable quantum computation via randomized compiling},\ }\href
	{https://doi.org/10.1103/PhysRevA.94.052325} {\bibfield  {journal} {\bibinfo
			{journal} {Phys. Rev. A}\ }\textbf {\bibinfo {volume} {94}},\ \bibinfo
		{pages} {052325} (\bibinfo {year} {2016})}\BibitemShut {NoStop}%
	\bibitem [{\citenamefont {Cai}\ and\ \citenamefont
		{Benjamin}(2019)}]{caiConstructingSmallerPauli2019}%
	\BibitemOpen
	\bibfield  {author} {\bibinfo {author} {\bibfnamefont {Z.}~\bibnamefont
			{Cai}}\ and\ \bibinfo {author} {\bibfnamefont {S.~C.}\ \bibnamefont
			{Benjamin}},\ }\bibfield  {title} {\bibinfo {title} {Constructing {{Smaller
					Pauli Twirling Sets}} for {{Arbitrary Error Channels}}},\ }\href
	{https://doi.org/10.1038/s41598-019-46722-7} {\bibfield  {journal} {\bibinfo
			{journal} {Sci Rep}\ }\textbf {\bibinfo {volume} {9}},\ \bibinfo {pages} {1}
		(\bibinfo {year} {2019})}\BibitemShut {NoStop}%
	\bibitem [{\citenamefont {Tsubouchi}\ \emph {et~al.}(2024)\citenamefont
		{Tsubouchi}, \citenamefont {Mitsuhashi}, \citenamefont {Sharma},\ and\
		\citenamefont {Yoshioka}}]{tsubouchi2024symmetric}%
	\BibitemOpen
	\bibfield  {author} {\bibinfo {author} {\bibfnamefont {K.}~\bibnamefont
			{Tsubouchi}}, \bibinfo {author} {\bibfnamefont {Y.}~\bibnamefont
			{Mitsuhashi}}, \bibinfo {author} {\bibfnamefont {K.}~\bibnamefont {Sharma}},\
		and\ \bibinfo {author} {\bibfnamefont {N.}~\bibnamefont {Yoshioka}},\
	}\bibfield  {title} {\bibinfo {title} {Symmetric clifford twirling for
			cost-optimal quantum error mitigation in early ftqc regime},\ }\href
	{https://arxiv.org/abs/2405.07720} {\bibfield  {journal} {\bibinfo  {journal}
			{arXiv preprint arXiv:2405.07720}\ } (\bibinfo {year} {2024})}\BibitemShut
	{NoStop}%
	\bibitem [{\citenamefont {Cai}\ \emph {et~al.}(2023)\citenamefont {Cai},
		\citenamefont {Babbush}, \citenamefont {Benjamin}, \citenamefont {Endo},
		\citenamefont {Huggins}, \citenamefont {Li}, \citenamefont {McClean},\ and\
		\citenamefont {O'Brien}}]{caiQuantumErrorMitigation2023}%
	\BibitemOpen
	\bibfield  {author} {\bibinfo {author} {\bibfnamefont {Z.}~\bibnamefont
			{Cai}}, \bibinfo {author} {\bibfnamefont {R.}~\bibnamefont {Babbush}},
		\bibinfo {author} {\bibfnamefont {S.~C.}\ \bibnamefont {Benjamin}}, \bibinfo
		{author} {\bibfnamefont {S.}~\bibnamefont {Endo}}, \bibinfo {author}
		{\bibfnamefont {W.~J.}\ \bibnamefont {Huggins}}, \bibinfo {author}
		{\bibfnamefont {Y.}~\bibnamefont {Li}}, \bibinfo {author} {\bibfnamefont
			{J.~R.}\ \bibnamefont {McClean}},\ and\ \bibinfo {author} {\bibfnamefont
			{T.~E.}\ \bibnamefont {O'Brien}},\ }\bibfield  {title} {\bibinfo {title}
		{Quantum error mitigation},\ }\href
	{https://doi.org/10.1103/RevModPhys.95.045005} {\bibfield  {journal}
		{\bibinfo  {journal} {Rev. Mod. Phys.}\ }\textbf {\bibinfo {volume} {95}},\
		\bibinfo {pages} {045005} (\bibinfo {year} {2023})}\BibitemShut {NoStop}%
	\bibitem [{\citenamefont {Yuan}\ \emph {et~al.}(2024)\citenamefont {Yuan},
		\citenamefont {Regula}, \citenamefont {Takagi},\ and\ \citenamefont
		{Gu}}]{yuan2024virtual}%
	\BibitemOpen
	\bibfield  {author} {\bibinfo {author} {\bibfnamefont {X.}~\bibnamefont
			{Yuan}}, \bibinfo {author} {\bibfnamefont {B.}~\bibnamefont {Regula}},
		\bibinfo {author} {\bibfnamefont {R.}~\bibnamefont {Takagi}},\ and\ \bibinfo
		{author} {\bibfnamefont {M.}~\bibnamefont {Gu}},\ }\bibfield  {title}
	{\bibinfo {title} {Virtual quantum resource distillation},\ }\href
	{https://doi.org/10.1103/PhysRevLett.132.050203} {\bibfield  {journal}
		{\bibinfo  {journal} {Phys. Rev. Lett.}\ }\textbf {\bibinfo {volume} {132}},\
		\bibinfo {pages} {050203} (\bibinfo {year} {2024})}\BibitemShut {NoStop}%
	\bibitem [{SM()}]{SM}%
	\BibitemOpen
	\href@noop {} {\bibinfo  {journal} {See the Supplementary Material for
			theoretical analysis, experimental details, and more data}\ }\BibitemShut
	{NoStop}%
	\bibitem [{\citenamefont {Patel}\ \emph {et~al.}(2016)\citenamefont {Patel},
		\citenamefont {Ho}, \citenamefont {Ferreyrol}, \citenamefont {Ralph},\ and\
		\citenamefont {Pryde}}]{raj2016fredkin}%
	\BibitemOpen
	\bibfield  {journal} {  }\bibfield  {author} {\bibinfo {author} {\bibfnamefont
			{R.~B.}\ \bibnamefont {Patel}}, \bibinfo {author} {\bibfnamefont
			{J.}~\bibnamefont {Ho}}, \bibinfo {author} {\bibfnamefont {F.}~\bibnamefont
			{Ferreyrol}}, \bibinfo {author} {\bibfnamefont {T.~C.}\ \bibnamefont
			{Ralph}},\ and\ \bibinfo {author} {\bibfnamefont {G.~J.}\ \bibnamefont
			{Pryde}},\ }\bibfield  {title} {\bibinfo {title} {A quantum fredkin gate},\
	}\href {https://doi.org/10.1126/sciadv.1501531} {\bibfield  {journal}
		{\bibinfo  {journal} {Science Advances}\ }\textbf {\bibinfo {volume} {2}},\
		\bibinfo {pages} {e1501531} (\bibinfo {year} {2016})}\BibitemShut {NoStop}%
	\bibitem [{\citenamefont {Ono}\ \emph {et~al.}(2017)\citenamefont {Ono},
		\citenamefont {Okamoto}, \citenamefont {Tanida}, \citenamefont {Hofmann},\
		and\ \citenamefont {Takeuchi}}]{Ono2017-ag}%
	\BibitemOpen
	\bibfield  {author} {\bibinfo {author} {\bibfnamefont {T.}~\bibnamefont
			{Ono}}, \bibinfo {author} {\bibfnamefont {R.}~\bibnamefont {Okamoto}},
		\bibinfo {author} {\bibfnamefont {M.}~\bibnamefont {Tanida}}, \bibinfo
		{author} {\bibfnamefont {H.~F.}\ \bibnamefont {Hofmann}},\ and\ \bibinfo
		{author} {\bibfnamefont {S.}~\bibnamefont {Takeuchi}},\ }\bibfield  {title}
	{\bibinfo {title} {Implementation of a quantum {controlled-SWAP} gate with
			photonic circuits},\ }\href {https://www.nature.com/articles/srep45353}
	{\bibfield  {journal} {\bibinfo  {journal} {Scientific Reports}\ }\textbf
		{\bibinfo {volume} {7}},\ \bibinfo {pages} {45353} (\bibinfo {year}
		{2017})}\BibitemShut {NoStop}%
	\bibitem [{\citenamefont {Peng}\ \emph {et~al.}(2025)\citenamefont {Peng},
		\citenamefont {Liu}, \citenamefont {Liu}, \citenamefont {Zhang},
		\citenamefont {Zhou},\ and\ \citenamefont {Lu}}]{peng2025fredkin}%
	\BibitemOpen
	\bibfield  {author} {\bibinfo {author} {\bibfnamefont {X.-J.}\ \bibnamefont
			{Peng}}, \bibinfo {author} {\bibfnamefont {Q.}~\bibnamefont {Liu}}, \bibinfo
		{author} {\bibfnamefont {L.}~\bibnamefont {Liu}}, \bibinfo {author}
		{\bibfnamefont {T.}~\bibnamefont {Zhang}}, \bibinfo {author} {\bibfnamefont
			{Y.}~\bibnamefont {Zhou}},\ and\ \bibinfo {author} {\bibfnamefont
			{H.}~\bibnamefont {Lu}},\ }\bibfield  {title} {\bibinfo {title} {Experimental
			shadow tomography beyond single-copy measurements},\ }\href
	{https://doi.org/10.1103/PhysRevApplied.23.014075} {\bibfield  {journal}
		{\bibinfo  {journal} {Phys. Rev. Appl.}\ }\textbf {\bibinfo {volume} {23}},\
		\bibinfo {pages} {014075} (\bibinfo {year} {2025})}\BibitemShut {NoStop}%
	\bibitem [{\citenamefont {Zhong}\ \emph {et~al.}(2018)\citenamefont {Zhong},
		\citenamefont {Li}, \citenamefont {Li}, \citenamefont {Peng}, \citenamefont
		{Su}, \citenamefont {Hu}, \citenamefont {He}, \citenamefont {Ding},
		\citenamefont {Zhang}, \citenamefont {Li}, \citenamefont {Zhang},
		\citenamefont {Wang}, \citenamefont {You}, \citenamefont {Wang},
		\citenamefont {Jiang}, \citenamefont {Li}, \citenamefont {Chen},
		\citenamefont {Liu}, \citenamefont {Lu},\ and\ \citenamefont
		{Pan}}]{PhysRevLett.121.250505}%
	\BibitemOpen
	\bibfield  {author} {\bibinfo {author} {\bibfnamefont {H.-S.}\ \bibnamefont
			{Zhong}}, \bibinfo {author} {\bibfnamefont {Y.}~\bibnamefont {Li}}, \bibinfo
		{author} {\bibfnamefont {W.}~\bibnamefont {Li}}, \bibinfo {author}
		{\bibfnamefont {L.-C.}\ \bibnamefont {Peng}}, \bibinfo {author}
		{\bibfnamefont {Z.-E.}\ \bibnamefont {Su}}, \bibinfo {author} {\bibfnamefont
			{Y.}~\bibnamefont {Hu}}, \bibinfo {author} {\bibfnamefont {Y.-M.}\
			\bibnamefont {He}}, \bibinfo {author} {\bibfnamefont {X.}~\bibnamefont
			{Ding}}, \bibinfo {author} {\bibfnamefont {W.}~\bibnamefont {Zhang}},
		\bibinfo {author} {\bibfnamefont {H.}~\bibnamefont {Li}}, \bibinfo {author}
		{\bibfnamefont {L.}~\bibnamefont {Zhang}}, \bibinfo {author} {\bibfnamefont
			{Z.}~\bibnamefont {Wang}}, \bibinfo {author} {\bibfnamefont {L.}~\bibnamefont
			{You}}, \bibinfo {author} {\bibfnamefont {X.-L.}\ \bibnamefont {Wang}},
		\bibinfo {author} {\bibfnamefont {X.}~\bibnamefont {Jiang}}, \bibinfo
		{author} {\bibfnamefont {L.}~\bibnamefont {Li}}, \bibinfo {author}
		{\bibfnamefont {Y.-A.}\ \bibnamefont {Chen}}, \bibinfo {author}
		{\bibfnamefont {N.-L.}\ \bibnamefont {Liu}}, \bibinfo {author} {\bibfnamefont
			{C.-Y.}\ \bibnamefont {Lu}},\ and\ \bibinfo {author} {\bibfnamefont {J.-W.}\
			\bibnamefont {Pan}},\ }\bibfield  {title} {\bibinfo {title} {12-photon
			entanglement and scalable scattershot boson sampling with optimal
			entangled-photon pairs from parametric down-conversion},\ }\href
	{https://doi.org/10.1103/PhysRevLett.121.250505} {\bibfield  {journal}
		{\bibinfo  {journal} {Phys. Rev. Lett.}\ }\textbf {\bibinfo {volume} {121}},\
		\bibinfo {pages} {250505} (\bibinfo {year} {2018})}\BibitemShut {NoStop}%
	\bibitem [{\citenamefont {Fiur\'a\ifmmode~\check{s}\else \v{s}\fi{}ek}\ and\
		\citenamefont {Hradil}(2001)}]{PhysRevA.63.020101}%
	\BibitemOpen
	\bibfield  {author} {\bibinfo {author} {\bibfnamefont {J.}~\bibnamefont
			{Fiur\'a\ifmmode~\check{s}\else \v{s}\fi{}ek}}\ and\ \bibinfo {author}
		{\bibfnamefont {Z.~c.~v.}\ \bibnamefont {Hradil}},\ }\bibfield  {title}
	{\bibinfo {title} {Maximum-likelihood estimation of quantum processes},\
	}\href {https://doi.org/10.1103/PhysRevA.63.020101} {\bibfield  {journal}
		{\bibinfo  {journal} {Phys. Rev. A}\ }\textbf {\bibinfo {volume} {63}},\
		\bibinfo {pages} {020101} (\bibinfo {year} {2001})}\BibitemShut {NoStop}%
	\bibitem [{\citenamefont {Nielsen}(2002)}]{nielsen2002simple}%
	\BibitemOpen
	\bibfield  {author} {\bibinfo {author} {\bibfnamefont {M.~A.}\ \bibnamefont
			{Nielsen}},\ }\bibfield  {title} {\bibinfo {title} {A simple formula for the
			average gate fidelity of a quantum dynamical operation},\ }\href
	{https://www.sciencedirect.com/science/article/pii/S0375960102012720}
	{\bibfield  {journal} {\bibinfo  {journal} {Physics Letters A}\ }\textbf
		{\bibinfo {volume} {303}},\ \bibinfo {pages} {249} (\bibinfo {year}
		{2002})}\BibitemShut {NoStop}%
	\bibitem [{\citenamefont {Yin}\ \emph {et~al.}(2017)\citenamefont {Yin},
		\citenamefont {Cao}, \citenamefont {Li}, \citenamefont {Liao}, \citenamefont
		{Zhang}, \citenamefont {Ren}, \citenamefont {Cai}, \citenamefont {Liu},
		\citenamefont {Li}, \citenamefont {Dai} \emph
		{et~al.}}]{doi:10.1126/science.aan3211}%
	\BibitemOpen
	\bibfield  {author} {\bibinfo {author} {\bibfnamefont {J.}~\bibnamefont
			{Yin}}, \bibinfo {author} {\bibfnamefont {Y.}~\bibnamefont {Cao}}, \bibinfo
		{author} {\bibfnamefont {Y.-H.}\ \bibnamefont {Li}}, \bibinfo {author}
		{\bibfnamefont {S.-K.}\ \bibnamefont {Liao}}, \bibinfo {author}
		{\bibfnamefont {L.}~\bibnamefont {Zhang}}, \bibinfo {author} {\bibfnamefont
			{J.-G.}\ \bibnamefont {Ren}}, \bibinfo {author} {\bibfnamefont {W.-Q.}\
			\bibnamefont {Cai}}, \bibinfo {author} {\bibfnamefont {W.-Y.}\ \bibnamefont
			{Liu}}, \bibinfo {author} {\bibfnamefont {B.}~\bibnamefont {Li}}, \bibinfo
		{author} {\bibfnamefont {H.}~\bibnamefont {Dai}}, \emph {et~al.},\ }\bibfield
	{title} {\bibinfo {title} {Satellite-based entanglement distribution over
			1200 kilometers},\ }\href {https://doi.org/10.1126/science.aan3211}
	{\bibfield  {journal} {\bibinfo  {journal} {Science}\ }\textbf {\bibinfo
			{volume} {356}},\ \bibinfo {pages} {1140} (\bibinfo {year}
		{2017})}\BibitemShut {NoStop}%
	\bibitem [{\citenamefont {G\"uhne}\ \emph {et~al.}(2007)\citenamefont
		{G\"uhne}, \citenamefont {Lu}, \citenamefont {Gao},\ and\ \citenamefont
		{Pan}}]{PhysRevA.76.030305}%
	\BibitemOpen
	\bibfield  {author} {\bibinfo {author} {\bibfnamefont {O.}~\bibnamefont
			{G\"uhne}}, \bibinfo {author} {\bibfnamefont {C.-Y.}\ \bibnamefont {Lu}},
		\bibinfo {author} {\bibfnamefont {W.-B.}\ \bibnamefont {Gao}},\ and\ \bibinfo
		{author} {\bibfnamefont {J.-W.}\ \bibnamefont {Pan}},\ }\bibfield  {title}
	{\bibinfo {title} {Toolbox for entanglement detection and fidelity
			estimation},\ }\href
	{https://journals.aps.org/pra/abstract/10.1103/PhysRevA.76.030305} {\bibfield
		{journal} {\bibinfo  {journal} {Phys. Rev. A}\ }\textbf {\bibinfo {volume}
			{76}},\ \bibinfo {pages} {030305} (\bibinfo {year} {2007})}\BibitemShut
	{NoStop}%
	\bibitem [{\citenamefont {Peres}(1996)}]{peres1996ppt}%
	\BibitemOpen
	\bibfield  {author} {\bibinfo {author} {\bibfnamefont {A.}~\bibnamefont
			{Peres}},\ }\bibfield  {title} {\bibinfo {title} {Separability criterion for
			density matrices},\ }\href {https://doi.org/10.1103/PhysRevLett.77.1413}
	{\bibfield  {journal} {\bibinfo  {journal} {Phys. Rev. Lett.}\ }\textbf
		{\bibinfo {volume} {77}},\ \bibinfo {pages} {1413} (\bibinfo {year}
		{1996})}\BibitemShut {NoStop}%
	\bibitem [{\citenamefont {Pan}\ \emph {et~al.}(2003)\citenamefont {Pan},
		\citenamefont {Gasparoni}, \citenamefont {Ursin}, \citenamefont {Weihs},\
		and\ \citenamefont {Zeilinger}}]{Pan2003purification}%
	\BibitemOpen
	\bibfield  {author} {\bibinfo {author} {\bibfnamefont {J.-W.}\ \bibnamefont
			{Pan}}, \bibinfo {author} {\bibfnamefont {S.}~\bibnamefont {Gasparoni}},
		\bibinfo {author} {\bibfnamefont {R.}~\bibnamefont {Ursin}}, \bibinfo
		{author} {\bibfnamefont {G.}~\bibnamefont {Weihs}},\ and\ \bibinfo {author}
		{\bibfnamefont {A.}~\bibnamefont {Zeilinger}},\ }\bibfield  {title} {\bibinfo
		{title} {Experimental entanglement purification of arbitrary unknown
			states},\ }\href {https://www.nature.com/articles/nature01623} {\bibfield
		{journal} {\bibinfo  {journal} {Nature}\ }\textbf {\bibinfo {volume} {423}},\
		\bibinfo {pages} {417} (\bibinfo {year} {2003})}\BibitemShut {NoStop}%
	\bibitem [{\citenamefont {Garcia-Escartin}\ and\ \citenamefont
		{Chamorro-Posada}(2013)}]{PhysRevA.87.052330}%
	\BibitemOpen
	\bibfield  {author} {\bibinfo {author} {\bibfnamefont {J.~C.}\ \bibnamefont
			{Garcia-Escartin}}\ and\ \bibinfo {author} {\bibfnamefont {P.}~\bibnamefont
			{Chamorro-Posada}},\ }\bibfield  {title} {\bibinfo {title} {swap test and
			hong-ou-mandel effect are equivalent},\ }\href
	{https://doi.org/10.1103/PhysRevA.87.052330} {\bibfield  {journal} {\bibinfo
			{journal} {Phys. Rev. A}\ }\textbf {\bibinfo {volume} {87}},\ \bibinfo
		{pages} {052330} (\bibinfo {year} {2013})}\BibitemShut {NoStop}%
	\bibitem [{\citenamefont {Abbott}\ \emph {et~al.}(2020)\citenamefont {Abbott},
		\citenamefont {Wechs}, \citenamefont {Horsman}, \citenamefont {Mhalla},\ and\
		\citenamefont {Branciard}}]{Abbott2020communication}%
	\BibitemOpen
	\bibfield  {author} {\bibinfo {author} {\bibfnamefont {A.~A.}\ \bibnamefont
			{Abbott}}, \bibinfo {author} {\bibfnamefont {J.}~\bibnamefont {Wechs}},
		\bibinfo {author} {\bibfnamefont {D.}~\bibnamefont {Horsman}}, \bibinfo
		{author} {\bibfnamefont {M.}~\bibnamefont {Mhalla}},\ and\ \bibinfo {author}
		{\bibfnamefont {C.}~\bibnamefont {Branciard}},\ }\bibfield  {title} {\bibinfo
		{title} {Communication through coherent control of quantum channels},\ }\href
	{https://doi.org/10.22331/q-2020-09-24-333} {\bibfield  {journal} {\bibinfo
			{journal} {{Quantum}}\ }\textbf {\bibinfo {volume} {4}},\ \bibinfo {pages}
		{333} (\bibinfo {year} {2020})}\BibitemShut {NoStop}%
	\bibitem [{\citenamefont {Rubino}\ \emph {et~al.}(2021)\citenamefont {Rubino},
		\citenamefont {Rozema}, \citenamefont {Ebler}, \citenamefont
		{Kristj\'ansson}, \citenamefont {Salek}, \citenamefont {Allard~Gu\'erin},
		\citenamefont {Abbott}, \citenamefont {Branciard}, \citenamefont {Brukner},
		\citenamefont {Chiribella},\ and\ \citenamefont
		{Walther}}]{rubino2021trajectories}%
	\BibitemOpen
	\bibfield  {author} {\bibinfo {author} {\bibfnamefont {G.}~\bibnamefont
			{Rubino}}, \bibinfo {author} {\bibfnamefont {L.~A.}\ \bibnamefont {Rozema}},
		\bibinfo {author} {\bibfnamefont {D.}~\bibnamefont {Ebler}}, \bibinfo
		{author} {\bibfnamefont {H.}~\bibnamefont {Kristj\'ansson}}, \bibinfo
		{author} {\bibfnamefont {S.}~\bibnamefont {Salek}}, \bibinfo {author}
		{\bibfnamefont {P.}~\bibnamefont {Allard~Gu\'erin}}, \bibinfo {author}
		{\bibfnamefont {A.~A.}\ \bibnamefont {Abbott}}, \bibinfo {author}
		{\bibfnamefont {C.}~\bibnamefont {Branciard}}, \bibinfo {author}
		{\bibfnamefont {i.~c.~v.}\ \bibnamefont {Brukner}}, \bibinfo {author}
		{\bibfnamefont {G.}~\bibnamefont {Chiribella}},\ and\ \bibinfo {author}
		{\bibfnamefont {P.}~\bibnamefont {Walther}},\ }\bibfield  {title} {\bibinfo
		{title} {Experimental quantum communication enhancement by superposing
			trajectories},\ }\href {https://doi.org/10.1103/PhysRevResearch.3.013093}
	{\bibfield  {journal} {\bibinfo  {journal} {Phys. Rev. Res.}\ }\textbf
		{\bibinfo {volume} {3}},\ \bibinfo {pages} {013093} (\bibinfo {year}
		{2021})}\BibitemShut {NoStop}%
\end{thebibliography}
\end{document}

\title{Supplementary Material for ``Experimental Quantum Channel Purification"}

\maketitle

\tableofcontents
\clearpage

\section{Theoritical details}

We analyze the circuit shown in Fig.2(a) in the main text.
Fredkin gate can be written as $\ketbra{0}{0}\otimes\mathbb{I}+\ketbra{1}{1}\otimes\mathbb{S}$, where the first qubit represents the control qubit and the identity operator $\mathbb{I}$ and the SWAP operator $\mathbb{S}$ act on the control and ancillary registers.
Then, the whole state in Fig.2(a) evolves as
\begin{equation}
\begin{aligned}
&\ketbra{+}\otimes\rho_{\mathrm{m}}\otimes\rho\\
&\xrightarrow{\mathrm{Fredkin}}\frac{1}{2}\left[\ketbra{0}\otimes(\rho_{\mathrm{m}}\otimes\rho)+\ketbra{1}\otimes(\rho\otimes\rho_{\mathrm{m}})+\ketbra{0}{1}\otimes(\rho_{\mathrm{m}}\otimes\rho)\mathbb{S}+\ketbra{1}{0}\otimes\mathbb{S}(\rho_{\mathrm{m}}\otimes\rho)\right]\\
&\xrightarrow{\mathcal{C}\otimes\mathcal{C}}\sum_{i,j}\frac{1}{2}p_ip_j[\ketbra{0}\otimes(E_i\rho_{\mathrm{m}}E_i^\dagger\otimes E_j\rho E_j^\dagger)+\ketbra{1}\otimes(E_i\rho E_i^\dagger\otimes E_j\rho_{\mathrm{m}}E_j^\dagger)+\ketbra{0}{1}\otimes(E_i\rho_{\mathrm{m}}E_j^\dagger\otimes E_j\rho E_i^\dagger)\mathbb{S}\\
&+\ketbra{1}{0}\otimes\mathbb{S}(E_j\rho_{\mathrm{m}}E_i^\dagger\otimes E_i\rho E_j^\dagger)]\\
&\xrightarrow{\mathrm{Fredkin}}\sum_{i,j}\frac{1}{2}p_ip_j[\ketbra{0}\otimes(E_i\rho_{\mathrm{m}}E_i^\dagger\otimes E_j\rho E_j^\dagger)+\ketbra{1}\otimes(E_j\rho_{\mathrm{m}}E_j^\dagger\otimes E_i\rho E_i^\dagger)+\ketbra{0}{1}\otimes(E_i\rho_{\mathrm{m}}E_j^\dagger\otimes E_j\rho E_i^\dagger)\\
&+\ketbra{1}{0}\otimes(E_j\rho_{\mathrm{m}}E_i^\dagger\otimes E_i\rho E_j^\dagger)]\\
&\xrightarrow{\text{measure control qubit and get $\ket{+}$}}\sum_{i,j}\frac{1}{4}p_ip_j\left(E_i\rho_{\mathrm{m}}E_i^\dagger\otimes E_j\rho E_j^\dagger+E_j\rho_{\mathrm{m}}E_j^\dagger\otimes E_i\rho E_i^\dagger+E_i\rho_{\mathrm{m}}E_j^\dagger\otimes E_j\rho E_i^\dagger+E_j\rho_{\mathrm{m}}E_i^\dagger\otimes E_i\rho E_j^\dagger\right)\\
&\xrightarrow{\text{discard the ancillary register}}\sum_{i,j}\frac{1}{4}p_ip_j[ E_j\rho E_j^\dagger+ E_i\rho E_i^\dagger+\delta_{ij} E_j\rho E_i^\dagger+\delta_{ij} E_i\rho E_j^\dagger]=\sum_{i}\frac{1}{2}(p_i+p_i^2)E_i\rho E_i^\dagger,
\end{aligned}
\end{equation}
where in the third line, we use the property of SWAP operator that $\mathbb{S}(E_i\otimes E_j)=(E_j\otimes E_i)\mathbb{S}$.
After normalization, the output state of the purified channel is $\mathcal{C}_{\mathrm{p}}^{+}(\rho)\equiv\mathcal{C}_{\mathrm{p}}(\rho)=\sum_ip_i\frac{1+p_i}{1+\sum_jp_j^2}E_i\rho E_i^\dagger$.

It can be similarly proved that, when the measurement result of the control qubit is $\ket{-}$, the output state is $\mathcal{C}_{\mathrm{p}}^{-}(\rho)=\sum_ip_i\frac{1-p_i}{1-\sum_jp_j^2}E_i\rho E_i^\dagger$.
Although the noise rate of channel $\mathcal{C}_{\mathrm{p}}^{-}$ is even worse than the original channel, the combination of $\mathcal{C}_{\mathrm{p}}^{+}$ and $\mathcal{C}_{\mathrm{p}}^{-}$ can result in a better virtual channel.
Specifically, when the measurement result of the control qubit is $\ket{+}$, one keeps the main register state as usually.
When the result is $\ket{-}$, instead of discarding the main register state, one can also keep it and classically add a minus sign to all measurement results obtained from this state.
Thus, one equivalently gets a virtual state, i.e. 
\begin{equation}
\mathcal{C}_{\mathrm{vp}}(\rho)=\frac{p_+\mathcal{C}_{\mathrm{p}}^{+}(\rho)-p_-\mathcal{C}_{\mathrm{p}}^{-}(\rho)}{p_+-p_-}=\sum_i\frac{p_i^2}{\sum_jp_j^2}E_i\rho E_i^\dagger.
\end{equation}

We now prove that the noise rates of $\mathcal{C}_{\mathrm{vp}}$ and $\mathcal{C}_{\mathrm{p}}$ are lower than the unpurified channel $\mathcal{C}$.
Suppose the dimension of the target quantum system is $d$, then there are totally $d^2$ Kraus operators with $p_0$ being the largest coefficient.
Then, we have 
\begin{equation}
\sum_ip_i^2=p_0\sum_{i}\frac{p_i}{p_0}p_i\le p_0\sum_{i}p_i=p_0.
\end{equation}
Therefore, $\frac{p_0^2}{\sum_jp_j^2}\ge p_0$ and the virtually purified channel $\mathcal{C}_{\mathrm{vp}}$ has a lower error rate compared with $\mathcal{C}$.
Similarly, as $p_0\frac{1+p_0}{1+\sum_jp_j^2}\ge p_0\frac{1+p_0}{1+p_0}=p_0$, we can prove that the purified channel $\mathcal{C}_{\mathrm{p}}$ also has a lower error rate compared with $\mathcal{C}$.

In our experiment, the control qubit is replaced by the Bell state $\ket{\Phi^+}=\frac{1}{\sqrt{2}}(\ket{00}+\ket{11})$ in Fig.2(b).
Following the similar derivation, one can change the measurement result $\ket{+}$ into $\ket{++}\&\ket{--}$ and change $\ket{-}$ into $\ket{+-}\&\ket{-+}$ to get the same purified channel.

\section{Experimental details}

\subsection{Initialization}

\begin{figure}[htbp!]
    \centering
    \includegraphics[width=0.5\linewidth]{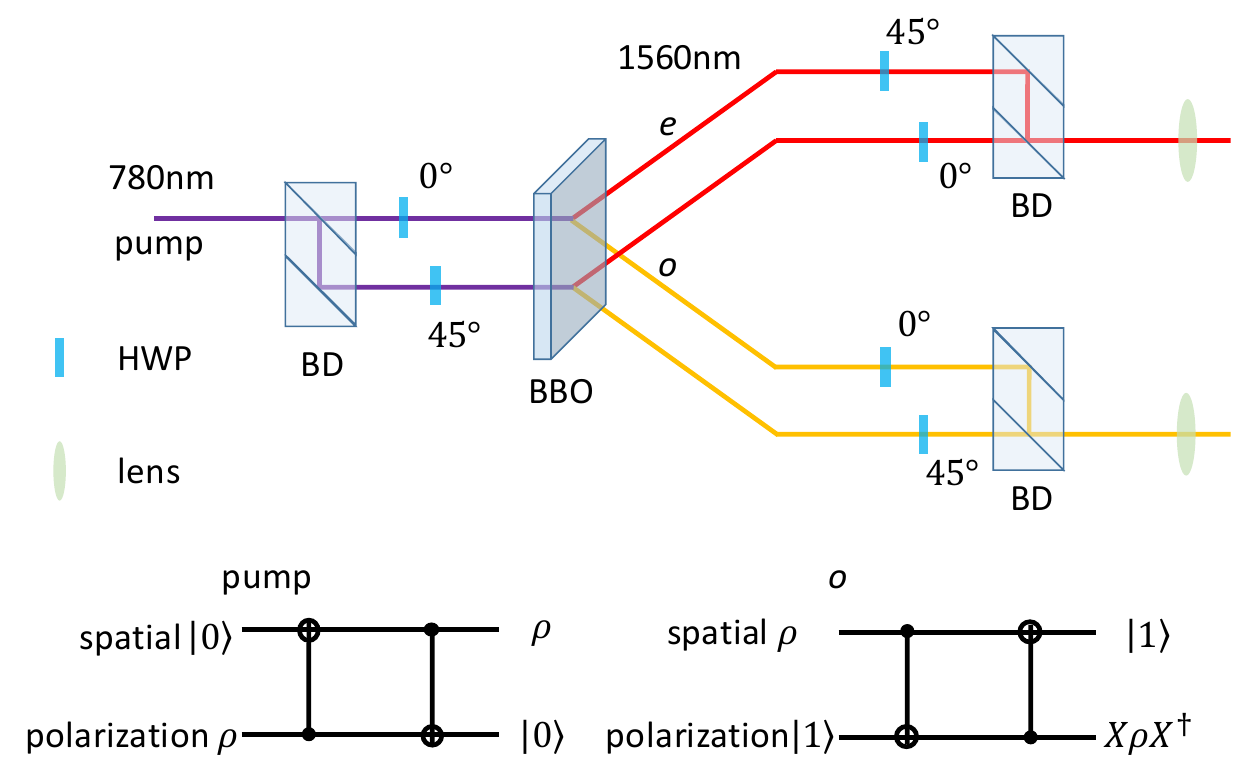}
    \caption{SPDC entanglement source.
   The two circuits below  are the spatial-polarization converter circuits.
            }
    \label{fig:detail}
\end{figure}

We use the interference-based beam-like spontaneous parametric down conversion (SPDC) entanglement source to prepare the entangled photon pairs.
The correlated photon pair can be generated from two isolated points in the barium borate (BBO) crystal, and they are entangled in spatial DoF.
To show other general EPR sources can also be used in our experiment, we introduce an extra step to convert this Einstein-Podolsky-Rosen (EPR) state into polarization DoF. %for more convenient manipulation.
We manage to use beam displacer (BD) to recombine the photon pair into a single spatial mode by inserting HWPs into different spatials.
As shown in Fig.~\ref{fig:detail}, the BD and HWPs can been seen as control-not (CNOT) gates and realize the exchange operation between two DoF.
After tilting these two BD and inserting the quartz crystal for precise spatial tuning and temporal compensation of the photon pairs, the Bell state in polarization modes can be prepared.
In the single-beam part between the two converters, the lens can be inserted for better coupling.
Moreover, multi-photon entanglement can be prepared by overlapping different EPR pairs on PBSs.
 
In this experiment, the pump laser is set with the central wavelength of 780 $\mathrm{nm}$, repetition rate of 76 $\mathrm{MHz}$, pulse duration of 150 $\mathrm{fs}$.
The power of pump laser focused on each of two points in the BBO crystal is set of 500 $\mathrm{mW}$ and the waist is 950 $\mathrm{um}$.
The full width at half maximum of the bandpass filters we use in front of the detectors is $\Delta\lambda = 30$ $\mathrm{nm}$.
As a result, the count rate of the EPR pair is about $50000$ $\text{s}^{-1}$ and the visibility of XX measurement result is $0.97$.
%count and fidelity

Following the entanglement source is the state preparation for the registers.
The Bell state $\ket{\Phi^+}$ for the spatial control registers are directly from the SPDC source after the convert operator, and the states in polarization DoF are reset to $\ket{\text{H}}$.
Then the different input states $\rho$ for the main register can be realizaed by rotating wave plates, and the preparation of the maximally-mixed state $\rho_{\mathrm{m}}$ for the ancillary register can be simply represented as
\begin{equation}
    \centering   
    \ketbra{0}\xrightarrow{\text{HWP}@22.5^{\circ}}\ketbra{+}{+}\xrightarrow{\text{dephasing by}\mathrm{YVO_4}}0.5\ketbra{0}{0}+0.5\ketbra{1}{1}.
\end{equation}

\subsection{Purification}

The controlled-SWAP operator has been explained in main text, here we focus on the Pauli channel in this part.
We use the QWP-HWP-QWP combination to realize the single-qubit rotation operator.
The rotation operator has following four components
\begin{equation}
    R_{\hat{n}}(\theta) = \mathrm{cos}(\frac{\theta}{2})I+\mathrm{sin}(\frac{\theta}{2})n_x(-iX)+\mathrm{sin}(\frac{\theta}{2})n_y(-iY)+\mathrm{sin}(\frac{\theta}{2})n_z(-iZ),  
\end{equation}
where $\hat{n}=(n_x,n_y,n_z)$ is the rotation axis and $\theta$ is the rotation angle.
There are various degree settings for the QWP-HWP-QWP combination to realize the operators.
The settings we used in the experiment are shown in the Table~\ref{tab:Pauli}.

\begin{table}[htbp!]
    \centering
    \begin{tabular}{c|c|c|c}
        \hline
        \hline
        \multirow{2}{*}{~\textbf{Operator}~}
        & \multicolumn{3}{c}{ \textbf{Settings} }\\
        \cline{2-4}
         &~~\textbf{QWP}~~&~~\textbf{HWP}~~&~~\textbf{QWP}~~\\
        \hline
        \hline

        $I$ & $0^{\circ}$ & $0^{\circ}$ & $0^{\circ}$ \\
        $-iX$ & $0^{\circ}$ & $45^{\circ}$ & $0^{\circ}$ \\
        $-iY$ & $0^{\circ}$ & $45^{\circ}$ & $90^{\circ}$ \\
        $-iZ$ & $0^{\circ}$ & $0^{\circ}$ & $90^{\circ}$ \\
        \hline
        \hline
    \end{tabular}
    \caption{Experimental setting of wave plates for different Pauli channels.}
    \label{tab:Pauli}
\end{table}

\subsection{Measurement}

\begin{figure}[htbp!]
    \centering
    \includegraphics[width=0.5\linewidth]{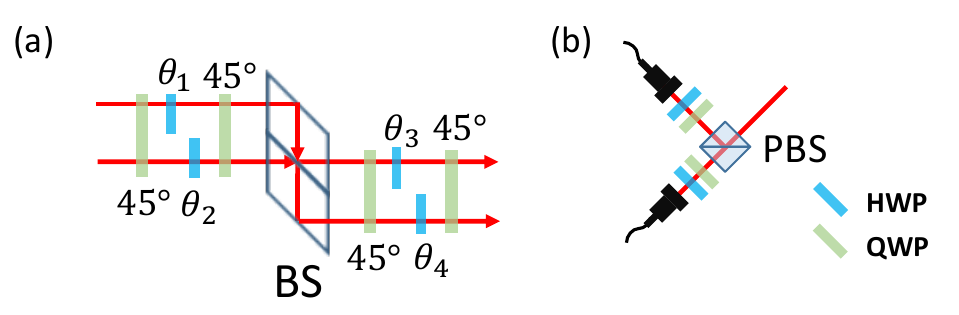}
    \caption{Details in measurement part.
    (a) Spatial Hadamand operator.
    (b) Polarization adjustment for SNSPD.
            }
    \label{fig:measurement}
\end{figure}

We use the beam splitter to realize $\textit{Hadamard}$ gate in spatial DoF.
The operation matrix of the BS can be given as 
\begin{equation}
    U_{\text{BS}}=e^{i\phi_0}\begin{bmatrix}
        \sqrt{R} e^{i\phi_\tau}&\sqrt{T} e^{i\phi_\rho}\\
        -\sqrt{T} e^{-i\phi_\rho}&\sqrt{R} e^{-i\phi_\tau}
    \end{bmatrix}.
\end{equation}

In this experiment, the  ratio of reflection and transmittance $R:T$ is $1:1$.
However, the phase shift are different to the photons $\ket{\text{H}}$ and $\ket{\text{V}}$, and they are $\left\{ \phi_{0\text{H}},\phi_{\tau \text{H}},\phi_{\rho\text{H}} \right\}$ and $\left\{\phi_{0\text{V}},\phi_{\tau \text{V}},\phi_{\rho\text{V}}\right\}$.
In order to implement polarization independent spatial \textit{Hadamard} gate, these different phase shift should be compensated.
Here we insert wave plates as phase tuners and tilt the BS to realize it, as shown in Fig.~\ref{fig:measurement}(a).

The phase tuner can be represented by Jones matrix in $\ket{\text{H}}/\ket{\text{V}}$ basis,
\begin{equation}
        \text{QWP}(\frac{\pi}{4})\text{HWP}(\theta)\text{QWP}(\frac{\pi}{4})=
        \begin{bmatrix}
        e^{-2i(\theta-\frac{\pi}{4})}&0\\
        0&e^{2i(\theta-\frac{\pi}{4})}
        \end{bmatrix}=
        \begin{bmatrix}
            e^{-i\Theta}&0\\0&e^{i\Theta}
        \end{bmatrix},
\end{equation}
is used to adjust the relative phase between photonic state $\ket{\text{H}}$ and $\ket{\text{V}}$.
While by tilting the BS, the delta phase is introduced between two spatials, as written in $\ket{0}/\ket{1}$ spatial basis
\begin{equation}
    \begin{bmatrix}
        1&0\\0&e^{i\Delta}
    \end{bmatrix}.
\end{equation}

Here, we take the input state $(\alpha\ket{\text{H}}+\beta\ket{\text{V}})\otimes(\gamma\ket{0}+\delta\ket{1})$ as the example.
The normalization coefficient is omitted.

\begin{equation}
    \begin{aligned}
        &\alpha\gamma\ket{\text{H}}\ket{0}+\beta\gamma\ket{\text{V}}\ket{0}+\alpha\delta\ket{\text{H}}\ket{1}+\beta\delta\ket{\text{V}}\ket{1}\\\xrightarrow{\mathrm{phase}}
        &\alpha\gamma e^{-i\Theta_1}\ket{\text{H}}\ket{u}+\beta\gamma e^{i\Theta_1}\ket{\text{V}}\ket{u}+\alpha\delta e^{i(\Delta-\Theta_2)}\ket{\text{H}}\ket{d}+\beta\delta e^{i(\Delta+\Theta_2)}\ket{\text{V}}\ket{d}\\\xrightarrow{\mathrm{BS}}
        &\alpha e^{i\phi_0\text{H}}[\gamma e^{i(\phi_{\tau\text{H}}-\Theta1)}-\delta e^{i(-\phi_{\rho\text{H}}+\Delta-\Theta_2)}]\ket{\text{H}}\ket{0}+\alpha e^{i\phi_0\text{H}}[\gamma e^{i(\phi_{\rho\text{H}}-\Theta1)}+\delta e^{i(-\phi_{\tau\text{H}}+\Delta-\Theta_2)}]\ket{\text{H}}\ket{1}\\+
        &\beta e^{i\phi_0\text{V}}[\gamma e^{i(\phi_{\tau\text{V}}+\Theta1)}-\delta e^{i(-\phi_{\rho \text{V}}+\Delta+\Theta_2)}]\ket{\text{V}}\ket{0}+\beta e^{i\phi_0\text{V}}[\gamma e^{i(\phi_{\rho\text{V}}+\Theta1)}+\delta e^{i(-\phi_{\tau\text{V}}+\Delta+\Theta_2)}]\ket{\text{V}}\ket{1}.
    \end{aligned}
    \label{eq:BS}
\end{equation}

In order to implement spatial X measurement, we require
\begin{equation}
\begin{aligned}
    &\Theta_1-\Theta_2+\Delta=\phi_{\rho\text{H}}+\phi_{\tau\text{H}}+\pi, \\
    &\Theta_1-\Theta_2-\Delta=-\phi_{\rho\text{V}}-\phi_{\tau\text{V}}+\pi.
\end{aligned}
\label{eq:r1}
\end{equation}

Back to Eq.~\eqref{eq:BS}

\begin{equation}
    \begin{aligned}
        &(\alpha e^{i(\phi_{0\text{H}}+\phi_{\tau\text{H}}-\Theta_1)}\ket{\text{H}}+\beta e^{i(\phi_{0\text{V}}+\phi_{\tau\text{V}}+\Theta_1)}\ket{\text{V}})(\gamma+\delta)\ket{0}\\+
        &(\alpha e^{i(\phi_{0\text{H}}+\phi_{\rho\text{H}}-\Theta_1)}\ket{\text{H}}+\beta e^{i(\phi_{0\text{V}}+\phi_{\rho\text{V}}+\Theta_1)}\ket{\text{V}})(\gamma-\delta)\ket{1}\\\xrightarrow{\mathrm{phase}}
        &(\alpha e^{i(\phi_{0\text{H}}+\phi_{\tau\text{H}}-\Theta_1-\Theta_3)}\ket{\text{H}}+\beta e^{i(\phi_{0\text{V}}+\phi_{\tau\text{V}}+\Theta_1+\Theta_3)}\ket{\text{V}})(\gamma+\delta)\ket{0}\\+
        &(\alpha e^{i(\phi_{0\text{H}}+\phi_{\rho\text{H}}-\Theta_1-\Theta_4)}\ket{\text{H}}+\beta e^{i(\phi_{0\text{V}}+\phi_{\rho\text{V}}+\Theta_1+\Theta_4)}\ket{\text{V}})(\gamma-\delta)\ket{1},
    \end{aligned}
\end{equation}

we need
\begin{equation}
    \begin{aligned}
        &2\Theta_3=\phi_{0\text{H}}+\phi_{\tau\text{H}}-\phi_{0\text{V}}-\phi_{\tau\text{V}}-2\Theta_1,\\
        &2\Theta_4=\phi_{0\text{H}}+\phi_{\rho\text{H}}-\phi_{0\text{V}}-\phi_{\rho\text{V}}-2\Theta_1.
    \end{aligned}
    \label{eq:r2}
\end{equation}

Above all, the spatial \textit{Hadamard} gate will be realized when Eq.~\eqref{eq:r1} and Eq.~\eqref{eq:r2} are satisfied, without affecting the photon's polarization.
If polarization measurement after the gate is unnecessary, i.e. trace operator in measurement 2 part, the compensate of $\Theta_3$ and $\Theta_4$ can be omitted.
If the input spatial state is $\ket{+}$, the photon will on the $\ket{0}$ path after the gate.
The visibility in this set is 0.950.

After measurement, each photon is collected by 4 superconduct nanowire single photon detectors (SNSPD), and the click event from each detector represents the measurement result in spatial and polarization DoF.
The detection efficiency of SNSPD is about $85\%$, but it is dependent on the polarization of the photon.
Because of this, we can not use single detector to collect photons in uncertain polarization in the trace operator.
To solve the problem, we divide the ancillary photon into orthogonal polarizations by PBS and collect it with two detectors.
Before each detector we insert a QWP and HWP to rotate the photon's polarization for the best efficiency.

\section{Entanglement distribution}\label{sec:Entanglement transmission}

\begin{figure*}[htbp!]
    \centering
    \includegraphics[width=1\linewidth]{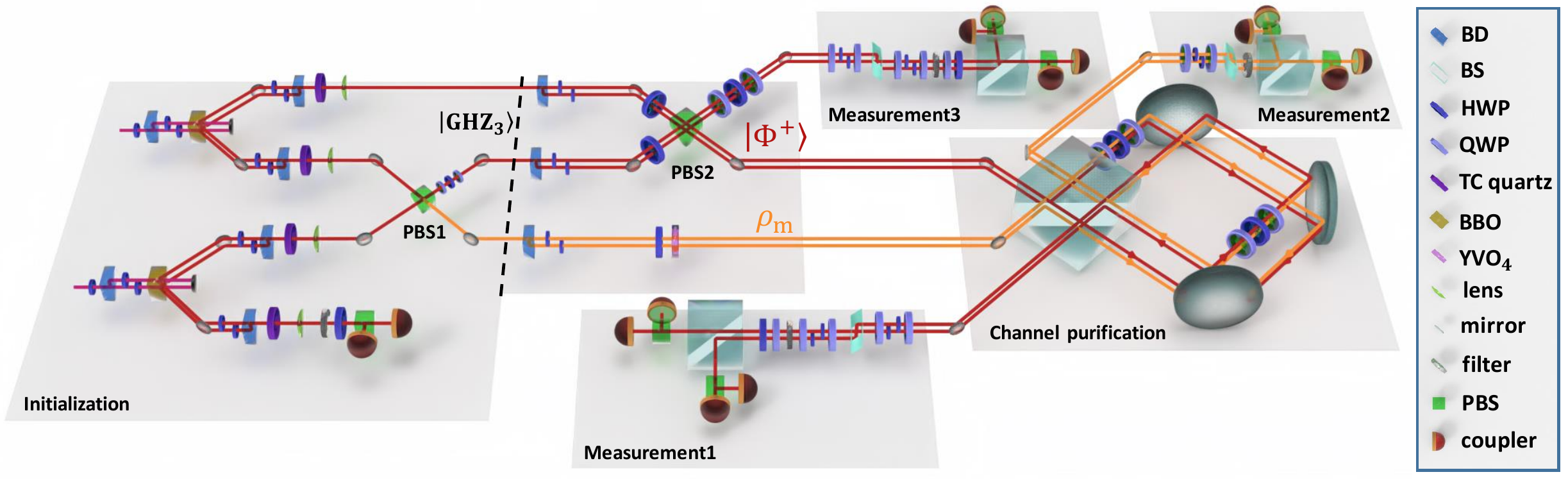}
    \caption{Experimental setup for entanglement transmission.
            }
    \label{fig:ED}
\end{figure*}

The detailed experimental setup for the demonstration of entanglement distribution is shown in Fig.~\ref{fig:ED}.
The channel purification and measurement part are the same as Fig.2(c) in the main text. 
Here we focus on the initialization part.
To prepare the distributed Bell state, two photons should be interfered on the PBS. This requires these two photons to be in the same layer.
Combining the requirement of the purification operation, these three photons should be entangled as the Greenberger-Horne-Zeilinger(GHZ) state in spatial DoF.
In our experiment, the GHZ state is prepared in polarization DoF by overlapping two photons in PBS1 from two different EPR source.
Then, the GHZ state is converted to spatial DoF with BDs and HWPs as shown in Fig.~\ref{fig:detail}.
Finally, the distributed Bell state $(\ket{00}+\ket{11})/\sqrt{2}$ is prepared by interfering two $\ket{+}$ photons on the PBS2.

In this experiment, all states in spatial DoF need to be measured in Pauli-$X$ basis.
The photon which is transmitted to Alice is entangled with others in spatial DoF and an extra local operator and classical communication (LOCC) operator is needed for Alice. 
However, this LOCC operator is only used in spatial DoF of photons, without affecting polarization DoF.
Also, this spatial measurement can be done before transmitting the photon to Alice.
%In this step, the LOCC operator are only used in spatial DoF of photons, without affecting polarization DoF.
If we project the spatial DoF of Alice's photon onto $\ket{+}$ before transmitting it, the channel purification protocol can be applied to Bob independently.
In this way, purification protocol targeting only on the most noisy channel independently has been realized.

\section{Tomographic data}
This section presents the reconstructed matrices, including the $\mathcal{X}$ matrix representation of the channels and the density matrix of the distributed states.
All matrices are calculated with the maximum likelihood method.
We use the reconstructed density matrix of the state after initial channels with $p$=1 instead of ideal Bell states to calculate the simulated line in Fig.4 in the main text.
To test the PPT criterion for the two initial states with $p=0.33$, we calculate the eigenvalues of their density matrices after partial transposition.
The results are $(0.03,0.28,0.33,0.36)$ and $(0.02,0.29,0.32,0.37)$, proving the two states are separable.

\begin{figure*}[htbp!]
    \centering
  \includegraphics[width=1\linewidth]{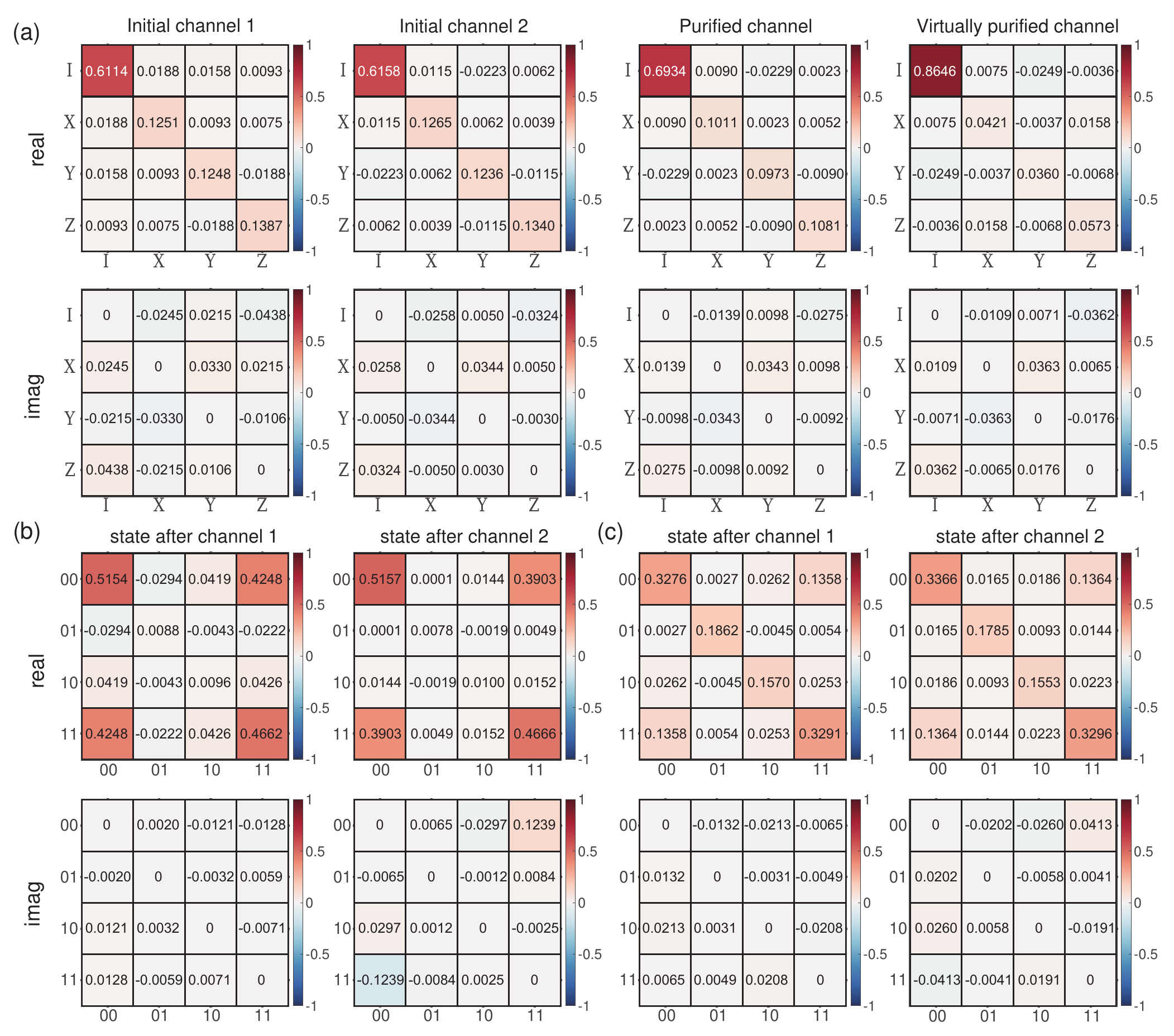}
    \caption{(a) Matrix representation of the channels with depolarizing noise with $p$=0.5.
    (b)(c) Reconstructed density matrix of the distributed state after depolarizing channels with $p$=1 and $p$=0.33.
    }
\end{figure*}

%\bibliographystyle{apsrev4-2}
%\bibliography{Ref_CP}